\documentclass[twocolumn,aps,prx,amsmath,amssymb,amsfonts,reprint,floatfix,superscriptaddress,longbibliography]{revtex4-2}

\usepackage{bm}
\usepackage{bbm}  
\usepackage{color}
\usepackage{enumerate}
\usepackage{graphicx}
\usepackage{microtype} 
\usepackage{multirow} 
\usepackage[normalem]{ulem}
\usepackage{xspace} 
\usepackage{float}
\usepackage{scalerel}
\usepackage{orcidlink}
\usepackage{cancel}

\usepackage{nameref} 

\hypersetup{colorlinks = true, urlcolor = blue, linkcolor = blue, citecolor = blue}

\allowdisplaybreaks 

\newcommand{\mb}[1]{\ensuremath{\mathbf{#1}}}
\newcommand{\mc}[1]{\ensuremath{\mathcal{#1}}}
\newcommand{\mr}[1]{\ensuremath{\mathrm{#1}}}
\newcommand{\mbb}[1]{\ensuremath{\mathbb{#1}}}
 
\newcommand{\tr}{\ensuremath{\mathrm{Tr}}}

\newcommand{\la}{\ensuremath{\langle}}
\newcommand{\ra}{\ensuremath{\rangle}}

\newcommand{\bra}[1]{\ensuremath{\langle #1 |}}
\newcommand{\ket}[1]{\ensuremath{| #1 \rangle}}

\newcommand{\ovl}[2]{\ensuremath{\langle #1 | #2 \rangle}}

\newcommand{\matel}[3]{\ensuremath{\langle #1 | #2 | #3 \rangle}}

\newcommand{\mi}{\ensuremath{\mathrm{i}}} 
\newcommand{\me}{\ensuremath{\mathrm{e}}} 

\newcommand{\Fex}{\ensuremath{\mathcal{F}_\mathrm{ex}}} 
\newcommand{\Fapx}{\ensuremath{\mathcal{F}_\mathrm{apx}}} 
\newcommand{\Fxeb}{\ensuremath{\mathcal{F}_\mathrm{\scriptscriptstyle{XEB}}}} 
\newcommand{\Fnxeb}{\ensuremath{\mathcal{F}_\mathrm{n\scriptscriptstyle{XEB}}}} 

\newcommand{\peps}{\ensuremath{\mathrm{\scriptscriptstyle{PEPS}}}} 
\newcommand{\mps}{\ensuremath{\mathrm{\scriptscriptstyle{MPS}}}} 

\newcommand{\D}{\ensuremath{\mathcal{D}}} 
\newcommand{\Dtr}{\ensuremath{\mathcal{D}_\mathrm{tr}}} 
\newcommand{\Dsat}{\ensuremath{\mathcal{D}_\mathrm{sat}}} 
\newcommand{\elayer}{\ensuremath{\varepsilon_\mathrm{layer}}} 
\newcommand{\ve}{\ensuremath{\varepsilon}} 

\newcommand{\CZ}{\ensuremath{\mathrm{CZ}}} 
\newcommand{\fSim}{\ensuremath{\mathrm{fSim}}} 
\newcommand{\fSimpp}{\ensuremath{\mathrm{fSim} (\pi/2, \pi/6)}} 

\newcommand{\Prob}{\ensuremath{\mathrm{Prob}}} 

\newcommand{\Eq}[1]{Eq.~\eqref{#1}}

\newcommand{\Fig}[1]{Fig.~\ref{#1}}

\newcommand{\Sec}[1]{Sec.~\ref{#1}}

\begin{document}

\title{Scalable projected entangled-pair state representation of random quantum circuit states}

\author{Sung-Bin B. Lee\,\orcidlink{0009-0009-2340-3682}}
\email{tomsb.lee@gmail.com}
\affiliation{Department of Physics and Astronomy, Seoul National University, Seoul 08826, Korea}
\author{Hee Ryang Choi\,\orcidlink{0009-0007-8069-1428}}
\affiliation{Department of Mechanical Engineering, Seoul National University, Seoul 08826, Korea}
\author{Daniel Donghyon Ohm\,\orcidlink{0009-0006-2204-2887}}
\affiliation{Department of Physics and Astronomy, Seoul National University, Seoul 08826, Korea}
\author{Seung-Sup B.~Lee\,\orcidlink{0000-0003-0715-5964}}
\email{sslee@snu.ac.kr}
\affiliation{Department of Physics and Astronomy, Seoul National University, Seoul 08826, Korea}
\affiliation{Center for Theoretical Physics, Seoul National University, Seoul 08826, Korea}
\affiliation{Institute for Data Innovation in Science, Seoul National University, Seoul 08826, Korea}

\date{\today}

\begin{abstract}
Classical simulation of a programmable quantum processor is crucial in identifying the threshold of a quantum advantage.
We demonstrate the simple update of projected entangled-pair states (PEPSs) in the Vidal gauge that represent random quantum circuit states, which center around recent quantum advantage claims.
Applied to square lattices of qubits akin to state-of-the-art superconducting processors,
the PEPS representation is exact for circuit depths less than $\Dtr$ = $\beta\log_2\chi$, where $\chi$ is the maximum bond dimension and $2 \lesssim \beta \lesssim 4$ depends on the choice of two-qubit gates, independent of the qubit number $n$.
We find the universal scaling behaviors of the state fidelity 
by treating large-scale circuits of $n \leq 10^{4}$, using $\chi \leq 128$ on a conventional CPU.
Our method has a polynomial scaling of computational costs with $n$ for circuit depth $\D=O(\log n)$ and is more advantageous than matrix product state approaches if $n$ is large.
This work underscores PEPSs as a scalable tool for benchmarking quantum algorithms with future potential for sampling applications using advanced contraction techniques.
\end{abstract}

\maketitle

\section{Introduction}

In recent years, the development of quantum devices has made significant progress in both the number of qubits and the fidelity of their operations.
It has fueled the expectation that for certain tasks designed to favor quantumness, quantum devices can achieve a computational advantage over their classical counterparts.

Sampling from a random quantum circuit (RQC)~\cite{Hangleiter2023} is extensively studied as such a task.
An RQC starts with the initial state of unentangled qubits, say $\ket{\mb{0}} = \ket{0}^{\otimes n}$, and entangles them through a unitary $U$ formed by randomly chosen single- and two-qubit gates.
Measurement of the RQC state $U \ket{\mb{0}}$ in the computational basis generates random bitstring samples, following Born's rule;
a bitstring $\mb{x} \in \{ 0, 1 \}^n$ is sampled with probability $p(\mb{x}) = | \matel{\mb{x}}{U}{\mb{0}} |^2$.
The question is whether a classical algorithm can generate bitstring samples efficiently, in a way indistinguishable from actual quantum measurements.
If the gate set is universal, an RQC of sufficiently large depth would draw a random state from the Haar ensemble, which typically has volume-law entanglement~\cite{Page1993,Foong1994,Sen1996,Eisert2010} and huge non-stabilizerness (the amount of non-Clifford resources, also called magic)~\cite{Bravyi2005, Howard2014, Howard2017, Seddon2021, Paviglianiti2025}.
A classical simulation of such states is thought to be computationally hard, as it requires exploring an exponentially large Hilbert space.

Despite complexity-theoretic proofs and arguments for an asymptotic quantum speedup for RQC sampling~\cite{Hangleiter2023}, the precise border of quantum advantage---in terms of the system size and operation fidelity---has been elusive. It is highlighted in a recent debate on Google LLC's ``quantum supremacy'' claim in 2019~\cite{Arute2019short}, where the Google team performed RQC sampling with the Sycamore processor, that consists of $n = 53$ superconducting qubits on a two-dimensional (2D) lattice.
Assuming that the linear cross-entropy benchmark (XEB), evaluated via bitstring sampling, is a proxy for fidelity, they estimated the fidelity $\mc{F} \simeq 0.2\%$ of the RQC output state for circuit depth $\mc{D} = 20$.
They further argued that classical algorithms cannot achieve the fidelity of a similar level, even using a state-of-the-art supercomputer for 10 000 years.

However, after a few years, this claim has been refuted by several studies using tensor networks.
In one approach~\cite{Pan2020,Pan2022,Pan2024,Gray2021,Zhao2024}, the sampling problem is recast as a contraction of a $(2+1)$-dimensional tensor network, where the two spatial dimensions encode the qubit layout, and the third dimension runs along the circuit depth. Using large-scale parallelization, Pan  \textit{et al.}~\cite{Pan2022a} demonstrated that the Sycamore experiment could be reproduced in 15 h on a cluster of 512 GPUs, while Liu \textit{et al.}~\cite{Liu2021} further reduced this time to 304 s on a supercomputer with 107 520 nodes (41 932 800 CPE cores).

Another approach~\cite{Zhou2020, Ayral2023, DeCross2024short,DeGirolamo2025} uses matrix product states (MPSs) that efficiently represent many-qubit states for which the bond dimension $\chi$ controls the trade-off between accuracy and cost.
Even for $\chi$ much smaller than the generic exact limit $2^{n/2}$, MPS representations are still accurate when the Schmidt coefficients (singular values) decay quickly, as shown by its widespread success in quantum many-body physics~\cite{Schollwoeck2011}.
Ayral \textit{et al.}~\cite{Ayral2023} simulated an RQC using the density-matrix renormalization group (DMRG) algorithm that variationally updates the constituent tensors of an MPS. 
Their closed simulation, which targets one bitstring at once rather than the full RQC state, achieved $\mc{F} > 0.2\%$ for the Sycamore circuit using a small bond dimension $\chi \simeq 100 \ll 2^{n/2}$ on conventional CPU machines.

Given further progress in the latest experiments~\cite{Choi2023, Morvan2024short, DeCross2024short, Gao2025short}, the arena of quantum-classical competition has moved to the region of larger $n$ and lower gate error.
The tensor network methods mentioned above, both the $(2+1)$-dimensional contraction and the DMRG, have complexity that scales exponentially with the size of the two-dimensional system, which calls for a different scheme.

Generally, the performance of a potentially useful quantum algorithm is associated with the fidelity of the states involved, while RQC sampling has found no practical applications yet.
Indeed, a sampling experiment achieving a high XEB score does not necessarily require a high-fidelity state and vice versa~\cite{Gao2024, Morvan2024short, Zhao2024}; even with a perfect quantum machine, one needs exponentially many measurements to determine bitstring probabilities or to reconstruct quantum states.
Since RQC is a quantum-efficient way to generate a classically hard state, a classical algorithm that can simulate large-scale RQC states with high fidelity would be transferrable to benchmarking useful quantum algorithms.

In this work, we demonstrate that a projected entangled-pair state (PEPS)~\cite{Nishino2001, Maeshima2001, Nishio2004, Verstraete2008, Orus2014, Orus2019, Cirac2021, McGinley2025} in the Vidal gauge~\cite{Vidal2003, Vidal2004, Vidal2007, Jiang2008, Kalis2012, Ran2012, Phien2015, Tindall2023, Puente2025, Lubasch2014, Lubasch2014a}, evolving with the simple update~\cite{Jiang2008}, can efficiently and faithfully represent the state of an RQC with local gates, over a depth $\D$ of interest in experiments.
As a concrete example, we apply our algorithm to a square lattice of qubits with single-qubit and nearest-neighbor two-qubit gates, reminiscent of the Sycamore experiment~\cite{Arute2019short}
(see \Sec{sec:random_quantum_circuit} and \Fig{fig:overview}).
The differences from Ref.~\cite{Arute2019short} are that our lattice is rotated by $45^{\circ}$, and the controlled-Z ($\CZ$) and two-qubit Haar random (2HR) gate sequences are further investigated here.
We emphasize that our lattice geometry is more challenging for a tensor network algorithm since the qubits on the boundaries have more connectivity, hence larger entanglement.

The PEPS representation has a perfect fidelity $\mc{F} = 1$ for $\D \leq \Dtr = \beta \log_2 \chi$, where $\Dtr$ is the depth after which PEPS tensors are truncated, $2 \lesssim \beta \lesssim 4$ depends on the type of two-qubit gates, and $\chi$ is the maximum bond dimension.
Unlike MPS-based approaches, the truncation depth $\Dtr$ does not depend on $n$.
While $\chi$ grows exponentially with the desired value of $\Dtr$, experimentally demonstrated depth $\D \simeq 20$ can be handled with a manageable size of $\chi$.

The Vidal gauge conditions provide a way to efficiently estimate the fidelity based on discarded weights~\cite{Sun2024}, without the need to contract two PEPS layers, which is computationally hard~\cite{Schuch2007, Markov2008, Guo2019}.
With this, we treat large RQCs up to $n = 10^4$ and identify the universal scaling behaviors of the fidelity using conventional CPU machines.
Since the simple update algorithm is highly parallelizable, our method can be applied to larger $n$, using a GPU cluster or a supercomputer.
(Note that we focus on achieving high-fidelity representation of an RQC state in this work; the issue of sampling, which requires the contraction of PEPS, will be addressed in a separate study.)

The remainder of this paper is organized as follows.
We describe our RQC setup and PEPS algorithm in  Secs.~\ref{sec:random_quantum_circuit} and \ref{sec:method},  respectively.
In \Sec{sec:fidelity_chi}, we discuss how the fidelity measures depend on $\D$, $\chi$, and the type of two-qubit gates.
In Sec.~\ref{sec:scalability}, we showcase a large-scale representation of RQC that reveals the universal scaling behavior of the fidelity.
We conclude with a discussion on the entanglement scaling laws and an outlook of our results in \Sec{sec:conclusion}.

\section{Setup: Random quantum circuit}
\label{sec:random_quantum_circuit}

\begin{figure}[t]
    \includegraphics[width=\linewidth]{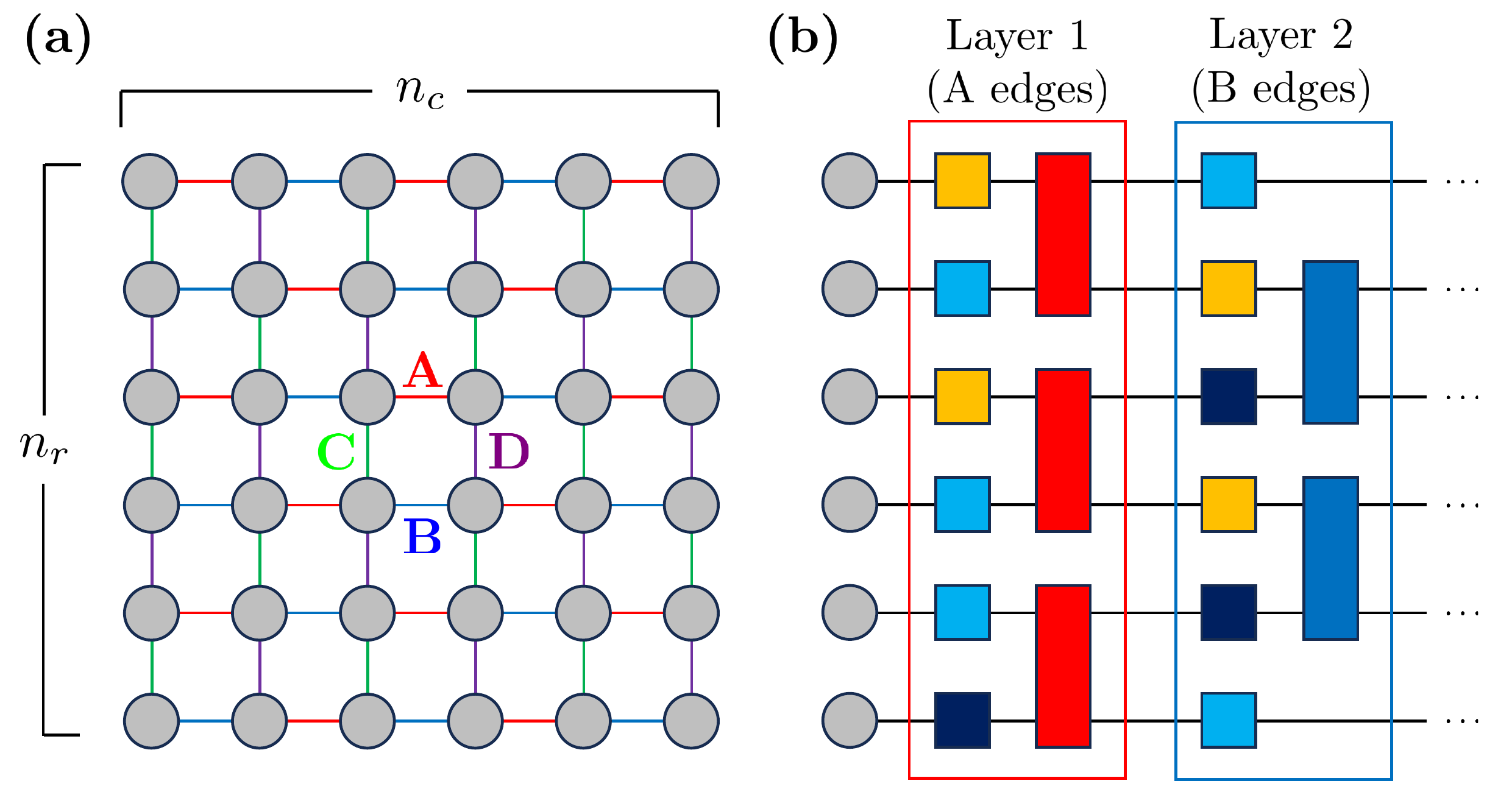}
    \caption{%
    (a) Layout of a square lattice of qubits (gray circles), whose state will be represented by a PEPS with the same lattice geometry (cf.~\Fig{fig:PEPS}).
    (b) At each layer of an RQC, all qubits are acted on by randomly chosen single-qubit gates (square boxes).
    Then, two-qubit gates (long rectangles) are applied to the qubit pairs selected by an alternating pattern (see \Sec{sec:random_quantum_circuit} for details).
    }
    \label{fig:overview}
    \vspace{-1em}
\end{figure}

We consider a square lattice of $n = n_r \times n_c$ qubits with $n_r $ rows and $n_c$ columns, as shown in \Fig{fig:overview}(a).
Nearest-neighbor qubits are connected by edges, which are grouped into four disjoint sets (A$-$D) distinguished by different colors.
The edges in each set form a brickwork pattern, ensuring that different edges do not share the same qubit.

Each circuit layer begins with the application of single-qubit gates to all qubits [see \Fig{fig:overview}(b)].
Every single-qubit gate is randomly drawn from three choices [depicted by square boxes in three different colors in \Fig{fig:overview}(b)]:
\begin{subequations}
\begin{align}
    \sqrt{X} &= \begin{pmatrix} 1 & -i \\ -i & 1 \end{pmatrix}/\sqrt{2} , \\
    \sqrt{Y} &= \begin{pmatrix} 1 & -1 \\ 1 & 1 \end{pmatrix}/\sqrt{2} , \\
    \sqrt{W} &= \begin{pmatrix} 1 & -\sqrt{i} \\ \sqrt{-i} & 1 \end{pmatrix}/\sqrt{2} , 
\end{align}
\label{eq:1QG}%
\end{subequations}
all represented in the basis of $\{ \ket{0}, \ket{1} \}$.
Note that \( \sqrt{X} \) and \( \sqrt{Y} \) are Clifford gates, while \( \sqrt{W} \) is non-Clifford.
Following the single-qubit gates, two-qubit gates are applied to the pairs of nearest-neighbor qubits connected by the edges belonging to an edge set [depicted by long rectangles in \Fig{fig:overview}(b), using the same color coding as the edges in \Fig{fig:overview}(a)].
Here we select the edge sets in a period-eight pattern of ABCD-CDAB-$\ldots$, which was used in the Sycamore experiment~\cite{Arute2019short}.
We denote the number of two-qubit gates applied up to depth $\D$ as $n_\mr{2qg}$, which scales as $n_\mr{2qg} (\D) \approx (2n - n_r - n_c) \D / 4$.

In later sections, we will compare three sequences that differ by the choice of two-qubit gates.
(1) In the $\CZ$ sequence, all the two-qubit gates are controlled-Z gates.
(2) The $\fSim(\theta,\phi)$ sequence uses
\begin{equation}
\fSim (\theta, \phi) = 
\begin{pmatrix}
1 & 0 & 0 & 0 \\
0 & \cos \theta & -\mi \sin \theta & 0 \\
0 & -\mi \sin \theta & \cos \theta & 0 \\
0 & 0 & 0 & \me^{-\mi \phi}
\end{pmatrix} ,
\label{eq:fSim_matrix}
\end{equation}
represented in the basis of $\{ \ket{00}, \ket{01}, \ket{10}, \ket{11} \}$.
The same values of $\theta \in [0, \pi/2]$ and $\phi \in [0, \pi]$ are used for all the two-qubit gates within the sequence.
Note that $\CZ$ is a special case of $\fSim$, i.e., $\CZ = \fSim(0,\pi)$, and $\fSim$ becomes non-Clifford for general $\theta$ and $\phi$.
(3) In the 2HR sequence, each two-qubit gate is independently drawn from the Haar random $\mr{U}(4)$ unitary matrices.

\vspace{2.5em}

\section{Method}
\label{sec:method}

In this section, we explain how we represent RQC states using PEPS in the Vidal gauge.
Section~\ref{sec:Vidal_gauge} provides the background on the Vidal gauge.
Readers who are experts in tensor networks may prefer to skip directly to \Sec{sec:algorithm} in which we describe our algorithm.
(On the other hand, readers unfamiliar with tensor networks may consult pedagogical reviews~\cite{Orus2014,Bruognolo2021}.)
In \Sec{sec:Fapx}, we develop a method for efficiently computing the approximate fidelity of the PEPS.

\begin{figure}
    \includegraphics[width=\linewidth]{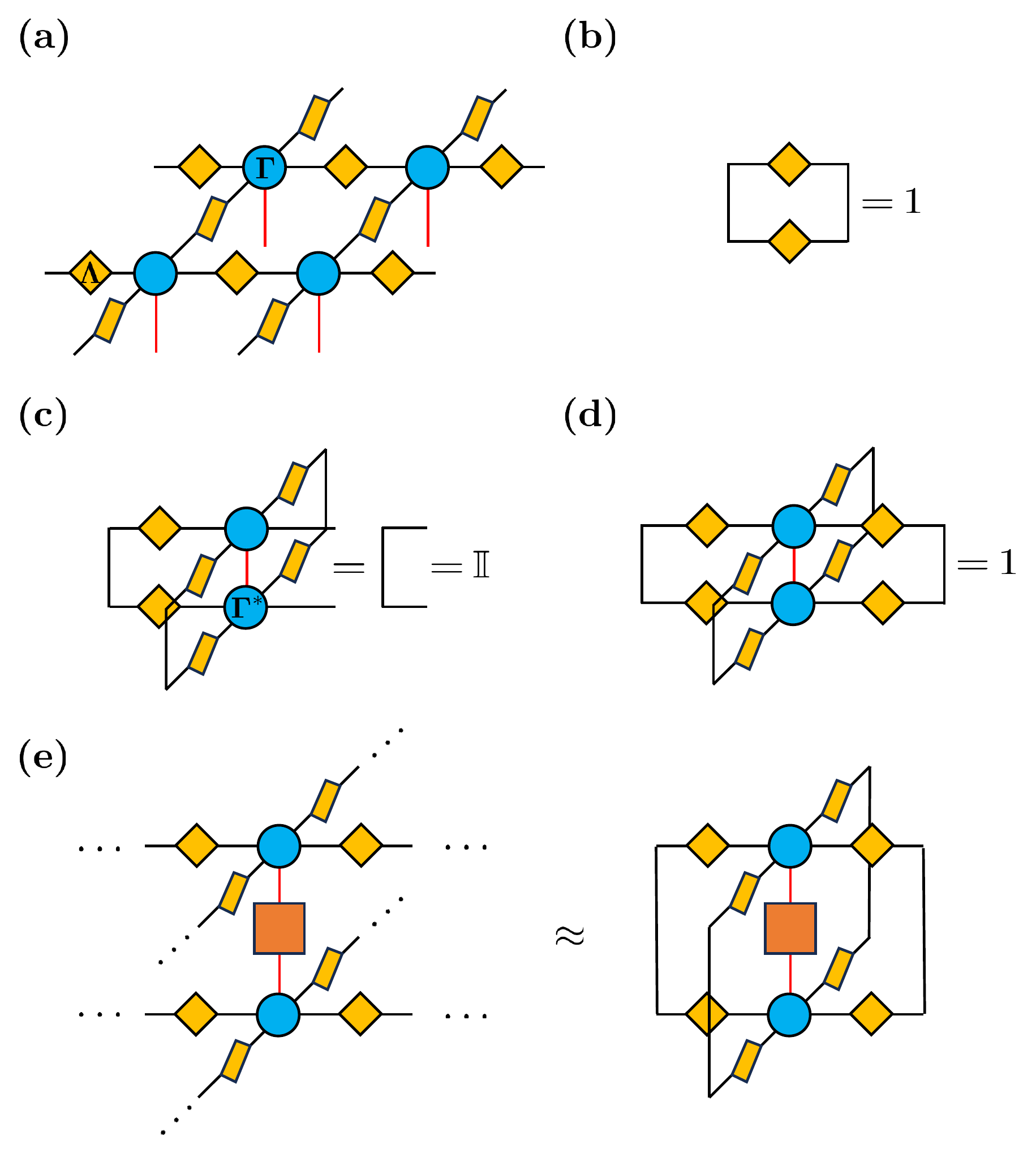}
    \caption{
(a) PEPS representation in the Vidal gauge for a ket state.
At every qubit site, there is a $\Gamma$ tensor (blue circle) surrounded by four $\Lambda$ tensors (yellow diamonds).
A physical leg (red line) of a $\Gamma$ tensor spans the space of a qubit with dimension $d = 2$.
The black lines connecting the tensors mean tensor contractions.
(b)--(d) Normalization conditions for the Vidal gauge, corresponding to Eqs.~\eqref{eq:Vidal_condition_Lambda}--\eqref{eq:Vidal_condition_Gamma_Lambda_2}, respectively.
The tensors in the bottom layer, which depicts a bra state, are complex conjugated.
(e) By approximating the environment as the nearest-neighbor $\Lambda$ tensors, a local observable (orange square) can be measured by contracting only the local part of a PEPS.
    }
    \label{fig:PEPS}
\end{figure}

\subsection{Vidal gauge}
\label{sec:Vidal_gauge}

A tensor network state on a tree graph, including MPS, can be brought into a canonical form, where the different bonds of the orthogonality center represent disjoint Hilbert spaces~\cite{Vidal2003,Vidal2004,Vidal2007,Schollwoeck2011}.
With the canonical form, the expectation values of local observables can be obtained by contracting only local sub-networks, avoiding the need to contract the full network.
In contrast, the graph structure of a PEPS reflects that of the problem to be solved (e.g., square-lattice geometry in this work) and thus contains loops, which makes the bond spaces correlated.
While it is not easy to treat a canonical form for such ``loopy'' PEPS (despite progress~\cite{Hyatt2019, Haghshenas2019, Zaletel2020, Soejima2020, Tepaske2021, Lin2022, Kadow2023, Sinha2024}), one can define a quasicanonical form, called the Vidal gauge~\cite{Tindall2023}, which is a higher-dimensional generalization of Vidal's $\Gamma$--$\Lambda$ form used in the infinite time-evolving block decimation approach for one-dimensional (1D) quantum systems~\cite{Vidal2007}.

Figure \ref{fig:PEPS}(a) shows a PEPS representation of qubits on a square lattice.
The $\Gamma$ tensors are degree-5 and complex, whose elements are denoted as $\Gamma^{(i)}_{x_i, \underline{1}, \underline{2}, \underline{3}, \underline{4}}$, where $i$ is the qubit site index, $x_i = 0, 1$ indicates the qubit state, and $\underline{1}, \dots, \underline{4}$ index the bond-space bases associated with the four legs of the $\Gamma^{(i)}$ tensor toward neighboring sites.
The $\Lambda$ tensors are degree-2 and positive diagonal, $\Lambda^{(i,j)}_{\underline{1},\underline{2}} = \sigma^{(i,j)}_{\underline{1}} \delta_{\underline{1}, \underline{2}}$, where $i$ and $j$ index the qubit sites connected by $\Lambda^{(i,j)}$.
We sort the singular values $\sigma^{(i,j)}_{\underline{1}}$ in descending order, $\sigma^{(i,j)}_{1} \geq \sigma^{(i,j)}_{2} \geq \sigma^{(i,j)}_{3} \geq \cdots > 0$.
These singular values contain the information of entanglement, though it is not precisely a bipartite entanglement because of the presence of loops.

Since we deal with finite-sized quantum systems, the PEPS has open boundaries.
The $\Gamma$ tensors on the boundaries have ``dummy'' legs with dimension $1$ toward the outside of the lattice. 
One can attach $\Lambda$ tensors, chosen to be scalar $1$s, to those dummy legs without changing the whole contraction of the PEPS.

The Vidal gauge means two normalization conditions.
First, the diagonal vector of every $\Lambda$ tensor is normalized,
\begin{equation}
\mr{Tr} [ \Lambda^{(i,j)} ]^2 = \sum_{\underline{1}} \left| \sigma^{(i,j)}_{\underline{1}} \right|^2 = 1,
\quad \forall i, j,
\label{eq:Vidal_condition_Lambda}
\end{equation}
which is depicted as a tensor network diagram in \Fig{fig:PEPS}(b).
Second, the contraction of a $\Gamma$ tensor, its neighboring $\Lambda$ tensors except one, and their complex conjugates [see \Fig{fig:PEPS}(c)] yields the identity,
\begin{equation}
\sum_{x_i, \underline{1}, \underline{2}, \underline{3}} \Gamma^{(i)}_{x_i,\underline{1}, \underline{2}, \underline{3}, \underline{4}} \Gamma^{(i)*}_{x_i,\underline{1}, \underline{2}, \underline{3}, \underline{5}} \left| \sigma^{(i,j)}_{\underline{1}} \sigma^{(i,k)}_{\underline{2}} \sigma^{(i,l)}_{\underline{3}} \right|^2 = \delta_{\underline{4}, \underline{5}},
\label{eq:Vidal_condition_Gamma_Lambda}
\end{equation}
which holds for all $i$'s and any choices of the neighbors $j, k, l$ to contract.
Combining Eqs.~\eqref{eq:Vidal_condition_Lambda} and \eqref{eq:Vidal_condition_Gamma_Lambda}, we derive another normalization condition,
\begin{equation}
\sum_{x_i, \underline{1}, \underline{2}, \underline{3}, \underline{4}} \Gamma^{(i)}_{x_i,\underline{1}, \underline{2}, \underline{3}, \underline{4}} \Gamma^{(i)*}_{x_i,\underline{1}, \underline{2}, \underline{3}, \underline{4}} \left| \sigma^{(i,j)}_{\underline{1}} \sigma^{(i,k)}_{\underline{2}} \sigma^{(i,l)}_{\underline{3}} \sigma^{(i,m)}_{\underline{4}} \right|^2 = 1,
\label{eq:Vidal_condition_Gamma_Lambda_2}
\end{equation}
which is visualized in \Fig{fig:PEPS}(d).

If the tensor network state were on a tree graph, the Vidal gauge means the exact canonical form.
Consider an environment, which is the same as the original network on a tree except that the $\Gamma^{(i)}$ and $\Lambda^{(i,j)}$ tensors sitting towards the nearest neighbor qubits $j$'s are blanked out.
The environment is a forest, i.e., a collection of disconnected trees, each of which has $\Gamma^{(j)}$ at the end.
Contracting all the legs of a tree state (containing $\Gamma^{(j)}$) and its complex conjugate, except for the legs toward site $i$, yields the identity.
It can be easily seen by applying the normalization condition \eqref{eq:Vidal_condition_Gamma_Lambda} [cf.~\Fig{fig:PEPS}(c)] from the farthest leaves of the tree from $\Gamma^{(j)}$.
Thus, when we measure local observables at the site $i$, the correlation between the site and the rest of the system is exactly captured  by the $\Lambda^{(i,j)}$ tensors.

On the other hand, when there are loops, the $\Lambda^{(i,j)}$ tensors surrounding $\Gamma^{(i)}$ are rather the approximate descriptions of the environment to $\Gamma^{(i)}$.
The simplest approximation for computing a local expectation value (e.g., Pauli-$Z$ measurement of qubit $i$, $\la Z_i \ra$) is given by approximating the loopy environment outside of site $i$ as uncorrelated $\Lambda$ tensors [see \Fig{fig:PEPS}(e)].
The time complexity of this measurement is $O(\chi^{z+1} zd)$, much cheaper than the explicit contraction of a double-layer tensor network.
This approximation is consistent with the simple update method we employ to evolve the tensors along an RQC (cf.~\Sec{sec:algorithm}); see \Sec{sec:local_and_global} for further discussion on the environment approximation.

\begin{figure}
   \includegraphics[width=\linewidth]{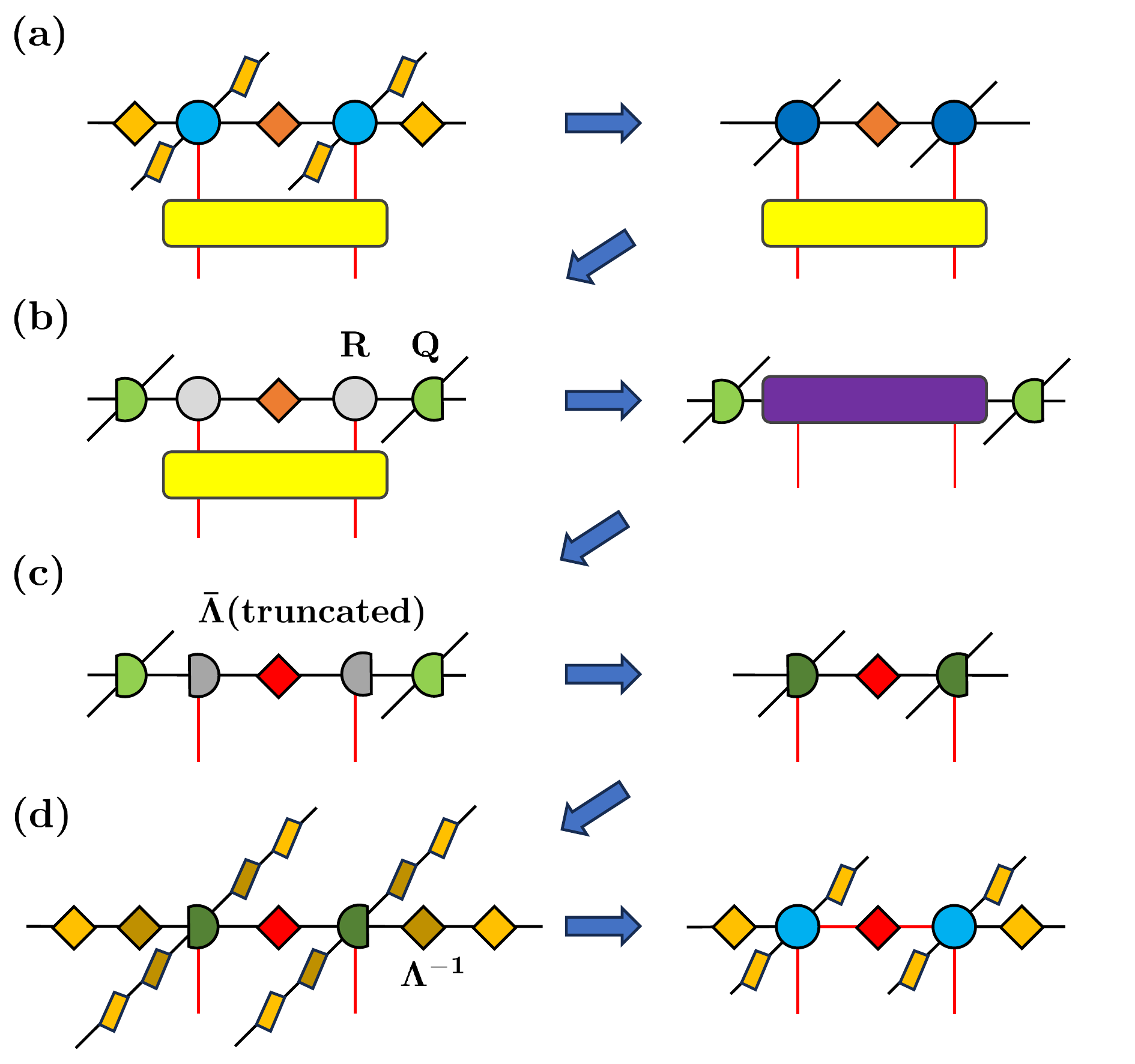}
    \caption{%
    Simple update method for updating the tensors associated with the nearest-neighbor pair of qubits, say $\Gamma^{(i)}$, $\Lambda^{(i,j)}$, and $\Gamma^{(j)}$.
    (a) Contract each $\Gamma$ tensor (light blue circle) with its neighboring $\Lambda$ tensors (yellow diamond), except for $\Lambda^{(i,j)}$ (orange diamond).
    (b) QR decompose the tensors on the qubit sites, which results in isometries $Q$ (light green objects) and degree-3 tensors $R$ (gray circle).
    This step, called bond projection \cite{Corboz2010, Li2012}, significantly reduces the computational cost in the later part.
    The bond projection itself has a time complexity of $O (\chi^{z+1} d^2)$, where $z = 4$ is the coordination number and $d = 2$ is the local dimension.
    Then, contract the single- and two-qubit gates with the $R$ tensors and $\Lambda^{(i,j)}$, leading to a degree-4 tensor (purple rectangle).
    If the simple update is used only for gauging [cf.~step (3) in \Sec{sec:algorithm}], we do not contract the gates.
    (c) Perform the truncated singular value decomposition (SVD) of the degree-4 tensor, which keeps at most $\chi$ largest singular values in the $\Lambda$ tensor (red diamond).
    Since finite singular values are lost, the diagonal vector of the $\Lambda$ tensor is normalized to restore the Vidal-gauge condition~\eqref{eq:Vidal_condition_Lambda}~\cite{Sun2024}.
    For the simple update gauging [cf.~step (3)], the bond dimension does not increase (due to the lack of gates); hence, it is not truncated.
    We contract the isometries (gray and light green objects) to get new pair of isometries (dark green objects).
    The computational cost of the SVD scales as $O(\chi^3 d^6)$, which is independent of the lattice geometry.
    (d) Factor out the neighboring $\Lambda$ tensors [which were absorbed in panel (a)] from the isometries by contracting their inverses (brown diamonds), which results in new $\Gamma^{(i)}$ and $\Gamma^{(j)}$.
    }
    \label{fig:simple_update}
\end{figure}

\subsection{Tensor network algorithm}
\label{sec:algorithm}

Below, we explain how we initialize and update a PEPS that represents an RQC state.
Once the circuit size and gate sequence are given, the only hyperparameter to set is the maximum bond dimension $\chi$, which controls the expressivity of the PEPS.

\begin{enumerate}[(1)]
\item
\textit{Initialization.}
The initial state of RQC is the product state, which we choose to be the all-$0$ state $\ket{0}$ without loss of generality.
Its PEPS representation has bond dimension $1$,
\begin{equation}
\Lambda^{(i,j)} = 1, \quad \Gamma^{(i)}_{x_i 1 \dots 1} = \delta_{x_i 0}, \quad \forall i, j ,
\label{eq:initial}
\end{equation}
which also satisfies the Vidal gauge conditions \eqref{eq:Vidal_condition_Lambda} and \eqref{eq:Vidal_condition_Gamma_Lambda}.
\end{enumerate}

For every circuit layer, we repeat steps (2) and (3) to evolve the PEPS.

\begin{enumerate}[(2)]
\item
\textit{Simple update with truncation.}
For every two-qubit gate, we update $\Gamma^{(i)}$, $\Lambda^{(i,j)}$, and $\Gamma^{(j)}$ associated with the target qubit pair $\la i, j \ra$, using the \textit{simple update} method~\cite{Jiang2008, Tindall2023} (see \Fig{fig:simple_update}).
(For qubits to which only single-qubit gates are applied, only the corresponding $\Gamma$ tensors are updated by contracting the degree-2 tensors representing the gates.)
Since two-qubit gates entangle qubits, the bond dimensions will grow exponentially $\sim 2^{\mc{D}/\beta}$ unless truncated, where $\mc{D}$ is the circuit depth and $\beta$ depends on the circuit and gate sequence (see Sec.~\ref{sec:Fex vs Fapx} for details).
To keep the bond dimensions below the prescribed maximum $\chi$, we keep the $\Gamma$ and $\Lambda$ tensors for bond indices $[1, \chi]$, associated with the largest singular values, and truncate the rest.
With this, the number of nonzero elements of $\Gamma$ and $\Lambda$ is upper bounded by $\chi^z d$ and $\chi$, respectively, throughout the computation.
\end{enumerate}

\begin{enumerate}[(3)]
\item
\textit{Simple update gauging.}
After step (2), the PEPS deviates from the Vidal-gauge normalization conditions since $\Gamma$ and $\Lambda$ tensors are updated.
To restore the normalization, we apply the simple update to all nearest-neighbor pairs of qubits without contracting gates [cf.~\Fig{fig:simple_update}], coined simple update gauging~\cite{Tindall2023}.
Note that this step does not involve truncations, so the fidelity does not decrease.

\end{enumerate}

In this work, we perform the gauging serially in the English-reading directions (horizontal pairs in the first row, left to right; vertical pairs between the first and second rows, left to right; horizontal pairs in the second row, left to right; and so on).
We find that only one or two sweeps of gauging suffice in achieving the normalization conditions within acceptable errors. 

\begin{figure*}
   \includegraphics[width=\textwidth]{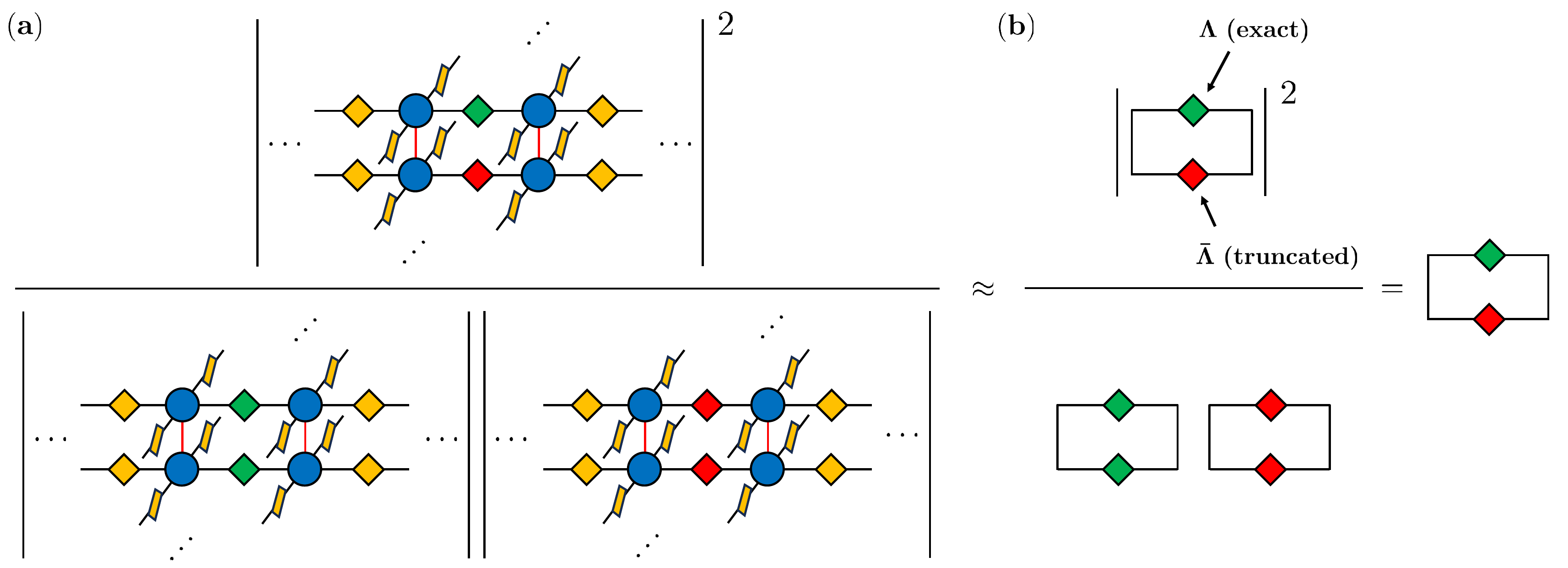}
    \caption{
    Tensor network diagrams depicting (a) the terms in \Eq{eq:f_t^ij} and (b) those in \Eq{eq:f_t^ij_approx}.
    }
    \label{fig:fidelity approximation}
\end{figure*}

The PEPS representation loses information when the tensors are truncated, which can happen only in step (2).
In other words, the PEPS is exact if $\chi$ is set large enough so that we keep all the tensor elements in the simple update.

The simple update we use here is the cheapest approach to evolve a PEPS since it involves only the local tensors by taking separate $\Lambda$ tensors as the proxy for the environment.
Due to the locality, the simple update can be applied to multiple sites simultaneously, which makes our algorithm highly parallelizable.
(Note that there is a newer, faster approach for performing the simple update based on belief propagation~\cite{Tindall2023}.)

Of course, the PEPS obtained by the simple update may not be the optimal representation for given bond dimensions since the environment is correlated, rather than being described by disconnected trees.
More elaborate approaches, such as the full update~\cite{Jordan2008}, can accurately describe the correlated environment, but their computational cost scales with a higher power of $\chi$, limiting the accessible range of $\chi$.
On the other hand, our loopy PEPS can contain loop correlations, which are unphysical correlations that do not influence physical observables but consume the expressivity.
It will be worthwhile to consider the methods that remove loop correlations, such as the graph-independent local truncations~\cite{Hauru2018, Lyu2021}, loop tensor network renormalization~\cite{Yang2017}, and nuclear norm regularization~\cite{Homma2024}, which is left for future studies.

\subsection{Computation of approximate fidelity}
\label{sec:Fapx}

A figure of merit in our study is the fidelity between the PEPS and the exact state,
\begin{equation}
\Fex (\mc{D},\chi)= 
\frac{ \big| \ovl{\psi_\mr{ex} (\mc{D})}{\psi_\peps (\mc{D},\chi)} \big|^2}{ \big\| \ket{\psi_\mr{ex} (\mc{D})} \big\|^2 \big\| \ket{\psi_\peps (\mc{D},\chi)} \big\|^2 },
\label{eq:Fex}
\end{equation}
where $\ket{\psi_\mr{ex} (\mc{D})}$ is the exact RQC state at depth $\mc{D}$ and $\ket{\psi_\peps (\mc{D},\chi)}$ is its PEPS approximation with the maximum bond dimension $\chi$.
Although $\ket{\psi_\mr{ex}}$ can be obtained by a brute-force state-vector simulation (the so-called Schr\"{o}dinger method) or as a truncation-free PEPS by setting a large $\chi$, it becomes exponentially hard for large qubit numbers $n$ or circuit depths $\mc{D}$; the state-vector simulation involves a vector of length $2^n$, and the truncation-free PEPS has bond dimensions exponentially increasing with $\D$.
Even when a PEPS form of $\ket{\psi_\mr{ex}}$ is available, computing the norm and the overlap requires the full contraction of two-dimensional tensor networks, which is computationally demanding.
Therefore, the exact fidelity $\Fex$ is not easy to verify~\cite{Schuch2007, Guo2019, Vanderstraeten2022}.

To circumvent this problem, we devise a method that efficiently estimates the fidelity, by generalizing an approach used in MPS simulations~\cite{Sun2024, Zhou2020} to the PEPS case.
The approximation is based on the observation that the fidelity decreases by truncating the tensors in the simple update, as discussed in \Sec{sec:algorithm}.
Assuming that the fidelity decreases at different circuit depths are independent, we can approximate the fidelity as a product~\cite{Ayral2023, Zhou2020}:
\begin{subequations}
\begin{align}
\Fex & \approx \prod_{t = 1}^{\mc{D}} \prod_{\la i, j \ra} f_t^{(i,j)}, \\
f_t^{(i,j)} &= \frac{ \big| \ovl{\psi_\peps^{(i,j);\mr{new}}}{\psi_\peps^{(i,j);\mr{old}}} \big|^2 }{ \big\| \ket{\psi_\peps^{(i,j);\mr{new}}} \big\|^2 \big\| \ket{\psi_\peps^{(i,j);\mr{old}}} \big\|^2 }, \label{eq:f_t^ij}
\end{align}
\end{subequations}
where $f_t^{(i,j)}$ is the fidelity between the states before and after the truncation in the simple update for a $(i,j)$ pair when considering the $t$th circuit layer.
The pretruncation PEPS $\ket{\psi_\peps^{(i,j);\mr{old}}}$ contains $\Gamma^{(i)}$, $\Lambda^{(i,j)}$, and $\Gamma^{(j)}$ that are obtained by the \textit{full} SVD.
The post-truncation PEPS $\ket{\psi_\peps^{(i,j);\mr{new}}}$ is equivalent to replacing $\Lambda^{(i,j)}$ from $\ket{\psi_\peps^{(i,j);\mr{old}}}$ with $\bar{\Lambda}^{(i,j)}$,
whose elements are given by
\begin{equation}
\bar{\Lambda}^{(i,j)}_{\underline{1}, \underline{2}} =
\begin{cases}
{\Lambda}^{(i,j)}_{\underline{1}, \underline{2}} = \sigma^{(i,j)}_{\underline{1}} \delta_{\underline{1}, \underline{2}}, & \underline{1}, \underline{2} \leq {\min} (\chi, \mr{rank} \, \Lambda^{(i,j)} ), \\
0, & \text{otherwise.}
\end{cases}
\label{eq:barLambda}
\end{equation}
The $0$s in $\bar{\Lambda}^{(i,j)}$ are multiplied with the elements of $\Gamma^{(i)}$ and $\Gamma^{(j)}$ associated with bond indices $> \chi$, reproducing the effect of the truncation.
(Of course, in actual calculations, we do not pad those $0$s to keep the memory usage small.)

Figure~\ref{fig:fidelity approximation}(a) shows the tensor network diagrams describing \Eq{eq:f_t^ij}.
We evaluate the individual terms on its right-hand side by invoking the approximation treating the environment as the identity, which underlies the simple update method.
With this, each term reduces to a loop contraction of two $\Lambda$ tensors [see \Fig{fig:fidelity approximation}(b)],
\begin{equation}
\begin{aligned}
f_t^{(i,j)} &\approx \frac{ \big| \mr{Tr} \big[ \Lambda^{(i,j)} \bar{\Lambda}^{(i,j)} \big] \big|^2 }{ \mr{Tr} \big[ (\Lambda^{(i,j)})^2 \big]  \,  \mr{Tr} \big[ (\bar{\Lambda}^{(i,j)})^2 \big] } \\
&= \mr{Tr} [ \Lambda^{(i,j)} \bar{\Lambda}^{(i,j)} ] =  \mr{Tr} [ (\bar{\Lambda}^{(i,j)})^2 ] = 1 - w^{(i,j)} ,
\end{aligned}
\label{eq:f_t^ij_approx}
\end{equation}
where we have used \Eq{eq:barLambda} and $\sigma^{(i,j)}_{\underline{1}} \in \mbb{R}$, and
\begin{equation}
w^{(i,j)}_t = \sum_{\underline{1} > \chi} \big| \sigma^{(i,j)}_{\underline{1}} \big|^2
\end{equation}
means the discarded weight due to the simple update.
Note that, after the truncation in the simple update, we divide the truncated $\Lambda$ tensor (without padded $0$s) by $\sqrt{\mr{Tr} [ (\bar{\Lambda}^{(i,j)})^2 ]} = \sqrt{1 - w^{(i,j)}_t}$ to restore the Vidal-gauge condition~\eqref{eq:Vidal_condition_Lambda}.

Therefore, we can compute the approximate fidelity,
\begin{equation}
\Fapx := \prod_{t = 1}^{\mc{D}} \prod_{\la i, j \ra} (1 - w^{(i,j)}_t) ,
\label{eq:Fapx}
\end{equation}
on the fly without extra computational overhead.
Despite the simplicity,  $\Fapx$ is a faithful approximation of the exact fidelity $\Fex$, as we will show in Sec.~\ref{sec:Fex vs Fapx} and \Fig{fig:exact_fid}.

\section{Fidelity for finite bond dimensions}
\label{sec:fidelity_chi}

The accuracy of a PEPS representation is quantified by the fidelity.
In this section, we discuss how the exact fidelity $\Fex$ [\Eq{eq:Fex}, \Sec{sec:Fex_chi}], the approximate fidelity $\Fapx$ [\Eq{eq:Fapx}, \Sec{sec:Fex vs Fapx}], and the normalized XEB (nXEB) [\Eq{eq:nXEB}, \Sec{sec:nXEB}] behave under finite bond dimensions.

\subsection{Exact fidelity}
\label{sec:Fex_chi}

\begin{figure*}
    \includegraphics[width=\linewidth]{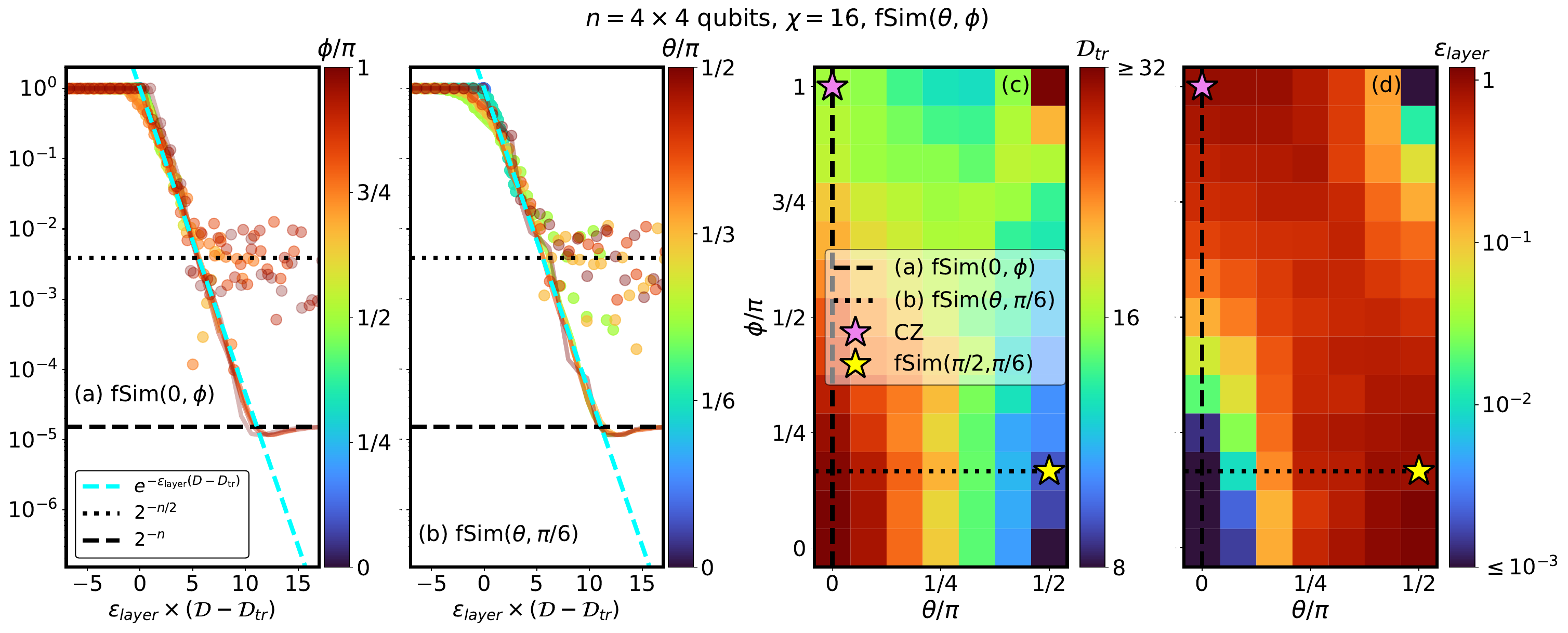}
    \caption{
    (a), (b) The exact fidelity $\Fex$ (solid lines) and the nXEB $\Fnxeb$ (dots, cf.~\Sec{sec:nXEB}) as a function of circuit depth $\D$ with varying the $\fSim$ gate parameters $(\theta,\phi)$, while the number of qubits $n$ and the maximum bond dimension $\chi$ are fixed.
    Panel (a) [panel (b)] shows the results for $\fSim(0,\phi)$, which includes $\CZ = \fSim(0,\pi)$ [$\fSim(\theta,\pi/6)$, which includes $\fSim(\pi/2,\pi/6)$ that we consider as the representative of the $\fSim$ gates for the rest of this paper].
    Different colors indicate different values of $\phi$ or $\theta$.
    The data of $\Fex (\D)$ and $\Fnxeb (\D)$ collapse on the scaling behaviors in Eqs.~\eqref{eq:Fex_D} and~\eqref{eq:Fapx_D}, respectively.
    (c), (d) The truncation depth $\Dtr$ and the error per layer $\elayer$ chosen to make the scaling collapse.
    For each $(\theta, \phi)$, we consider one circuit instance and the same values of $\Dtr$ and $\elayer$ are used to rescale both $\Fex (\D)$ and $\Fxeb (\D)$.
    }
   \label{fig:angles}
   \vspace{-1em}
\end{figure*}

Based on our numerical results, we find that $\Fex$ manifests a three-stage dependence on $\D$,
\begin{equation}
\Fex (\mc{D}) \simeq
\begin{cases}
1, & \mc{D} < \Dtr, \\
\me^{- \elayer (\mc{D} - \Dtr)}, & \Dtr \leq \mc{D} < \Dsat, \\
2^{-n}, & \Dsat \leq \mc{D} ,
\end{cases}
\label{eq:Fex_D}
\end{equation}
where the truncation depth $\Dtr$, the error per circuit layer $\elayer$, and the saturation depth $\Dsat$ depend on the number of qubits $n$, the gate sequence, and $\chi$.
In the shallow-depth regime, $\D < \Dtr$, a PEPS is exact since the bond dimensions are not larger than the maximum $\chi$ and thus the tensors are not truncated (cf.~\Sec{sec:algorithm}).
Since the bond dimensions increase exponentially, one encounters bonds with dimensions larger than $\chi$ at $\D = \Dtr$.
Once truncations are made, $\Fex$ starts to decrease exponentially; 
in other words, it decays by a factor $\me^{-\elayer} < 1$ after every layer.
The decay stops when $\Fex$ reaches the saturation limit,
\begin{equation}
\Fex \simeq
\mathop{\mbb{E}}_{\ket{\psi} \sim H}
\left[
\frac{ \big| \ovl{\psi_\mr{ex}}{\psi} \big|^2 }{ \big\| \ket{\psi_\mr{ex}} \big\|^2 \big\| \ket{\psi} \big\|^2}
\right]= 2^{-n} ,
\end{equation}
which is the average fidelity over the Haar ensemble of $n$-qubit states, with respect to the exact state $\ket{\psi_\mr{ex}}$. 
Therefore, for $\D \geq \Dsat = \Dtr + (n/\elayer) \ln 2$, the PEPS representation is not better than guessing a maximally mixed state with fidelity $2^{-n}$. While detailed parameters may differ, the general scheme of fidelity decay is universal across different gate sets~\cite{Ware2023}.

As a concrete example, we consider the $\fSim$ sequences on $n = 4 \times 4$ qubits that are small enough to compute $\ket{\psi_\mr{ex}}$ and $\Fex$ explicitly, using the exact state-vector simulation.
In \Fig{fig:angles}, we show the result for fixed $n$ and $\chi$, with changing the $\fSim$ gate parameters $(\theta, \phi)$.
Solid lines in Figs.~\ref{fig:angles}(a) and~\ref{fig:angles}(b) agree with \Eq{eq:Fex_D}.
We estimate $\Dtr$ and $\elayer$ for various $\theta$ and $\phi$, by fitting the numerical result of $\Fex (\D)$ to \Eq{eq:Fex_D}, to study their gate dependence.

\begin{figure*}
    \includegraphics[width=\linewidth]{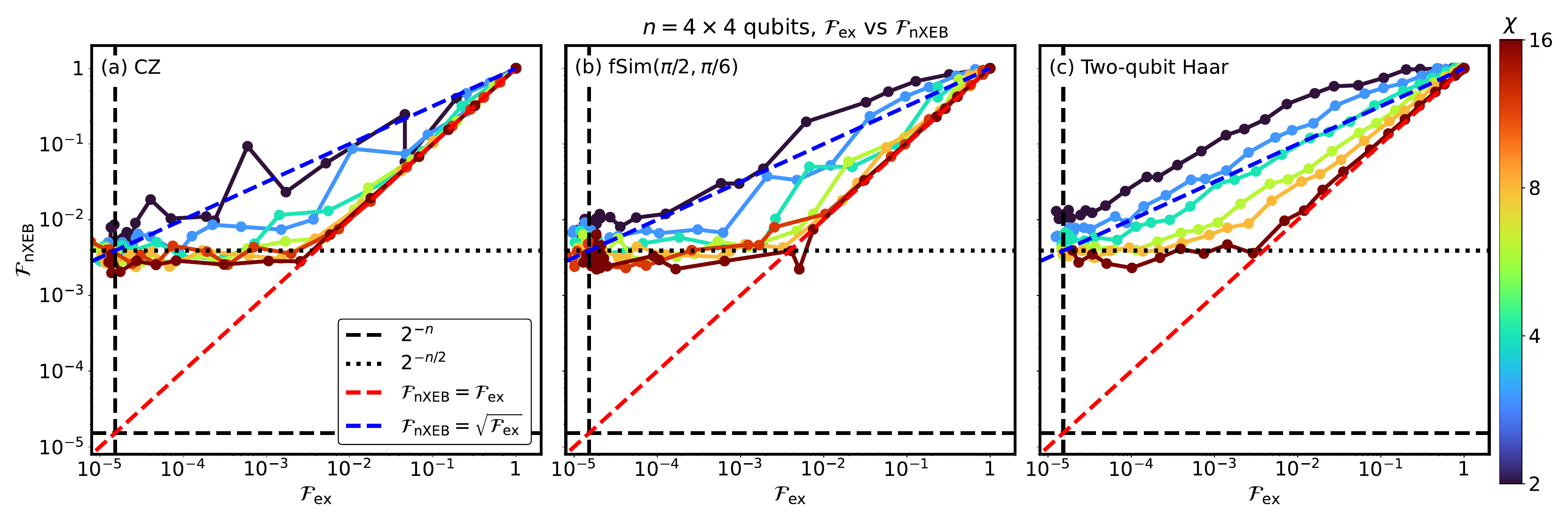}
    \caption{
    The nXEB $\Fnxeb$ vs the exact fidelity $\Fex$ for three different choices of two-qubit gates, (a) $\CZ$; (b) $\fSim (\pi/2, \pi/6)$; and (c) 2HR.
    Each dot marks $(\la \Fex \ra, \la \Fnxeb \ra)$ for a certain depth $\D$ and bond dimension $\chi$, where $\la \cdot \ra$ means the average over ten circuit instances.
    The dots for the same $\chi$ are in the same color and connected by lines.
    }
   \label{fig:XEB vs fidelity}
\end{figure*}

\begin{figure*}
    \includegraphics[width=\linewidth]{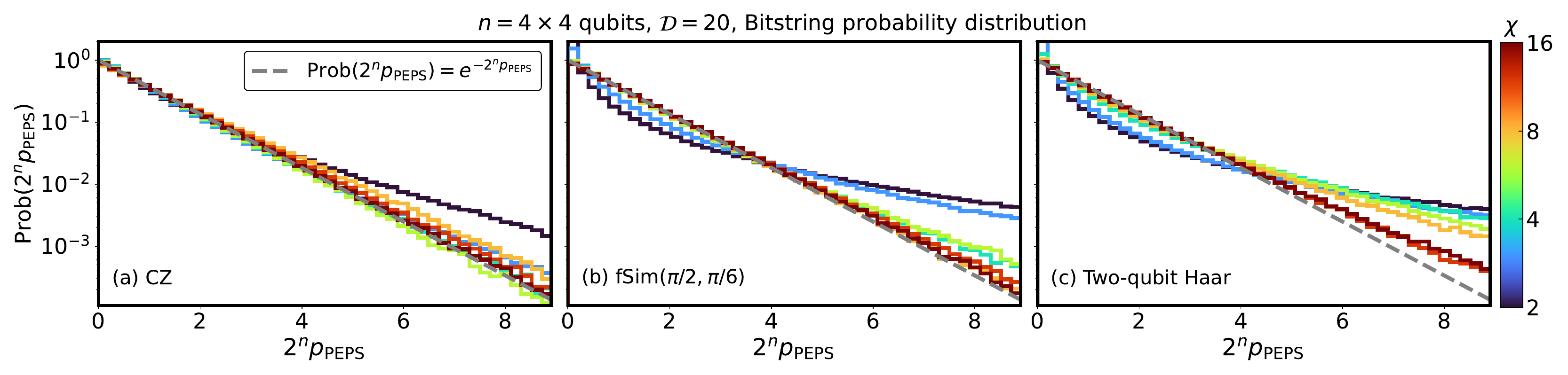}
    \caption{
    The distribution of bitstring probabilities $p_\peps$ from the PEPS representation of RQCs with (a) $\CZ$, (b) $\fSimpp$, and (c) 2HR gates.
    Different line colors indicate different $\chi$.
    Gray dashed lines represent the PTD, $\Prob (p) = 2^n \mr{exp} (-2^n p)$.
    Here, we choose $\D = 20$ at which exact state-vector simulations of these RQCs result in the PTD; that is, the deviation of the plotted curves from the PTD line is due to finite bond dimensions.}
   \label{fig:PTD}
\end{figure*}

As shown in Figs.~\ref{fig:angles}(c) and~\ref{fig:angles}(d), $\Dtr$ and $\elayer$ are anticorrelated.
The gate with larger $\Dtr$ and smaller $\elayer$ is easier to handle with PEPS since the fidelity can be kept larger for given $\D < \Dsat$ and $\chi$.
The easiest cases are $(\theta,\phi) = (0,0)$ and $(\pi/2,\pi)$ in which $\Dtr = \infty$ and $\elayer = 0$.
While $\fSim(0,0)$ is the identity and does not generate entanglement, $\fSim(\pi/2,\pi)$ can increase entanglement in general, since it is a SWAP gate preceded by local phase shifts,
\begin{subequations}
\begin{equation}
\fSim (\pi / 2, \pi) = \mr{SWAP} \cdot [ P(-\pi/2) \otimes P (-\pi/2) ],
\end{equation}
where
\begin{equation}
\mr{SWAP} = \begin{pmatrix}
        1 & 0 & 0 & 0 \\
        0 & 0 & 1 & 0 \\
        0 & 1 & 0 & 0 \\
        0 & 0 & 0 & 1 \\
    \end{pmatrix},
\quad
P (-\pi/2) = \begin{pmatrix}
        1 & 0 \\
        0 & -i
    \end{pmatrix}.
\end{equation}%
\end{subequations}%
Note that the operator entanglement entropy of $\fSim(\pi/2,\pi)$ is maximal for two-qubit gates, i.e., $\log 4$ (cf.~\Sec{sec:Fex vs Fapx}).
However, the initial state of an RQC is a product state $\ket{0}$, so SWAP as the only multiqubit gate in the circuit keeps the RQC state being a product state (see~\Sec{sec:random_quantum_circuit}).
On the other hand, the hardest gate to simulate is $\fSim(\pi/2,0)$, where $\Dtr$ is the smallest and $\elayer$ is the largest.
Recently, this choice was also underscored as the hardest gate for classical simulations in a different sense~\cite{Gao2024}; the effectiveness of spoofing XEB is minimized for the $\fSim(\pi/2,0)$ sequence.

\begin{figure*}
    \centering
    \includegraphics[width=\textwidth]{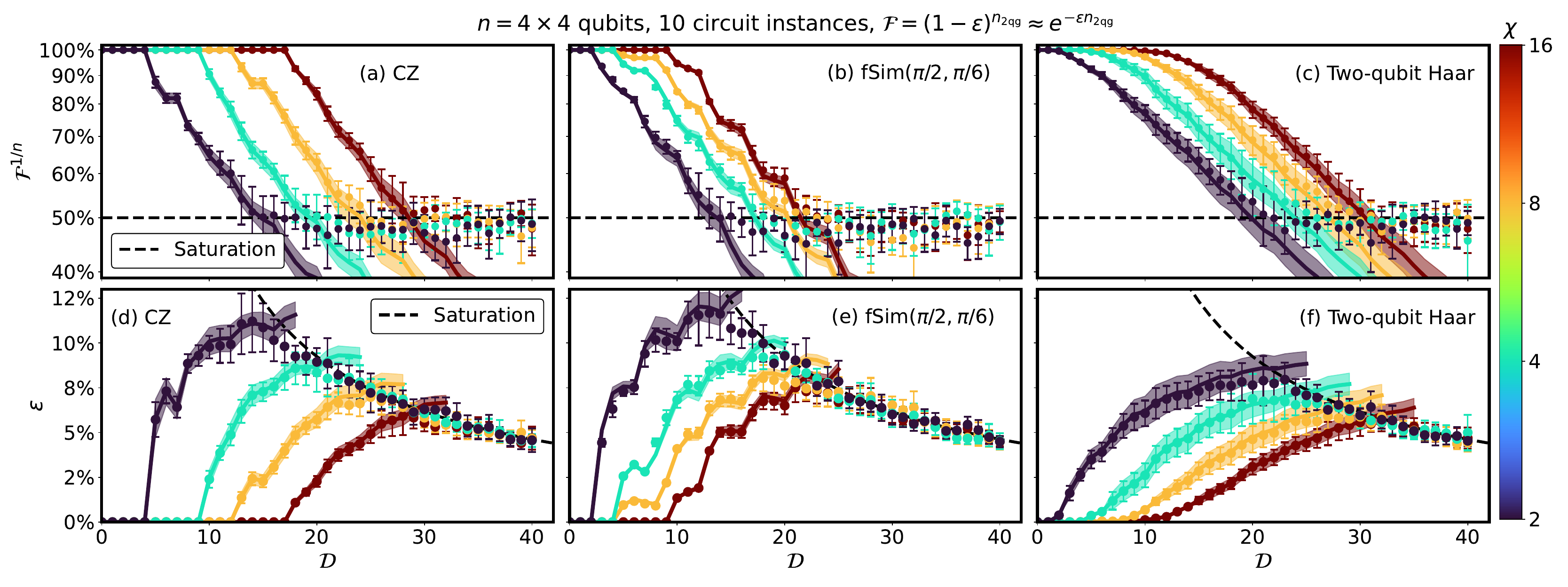}
    \caption{
      Comparison between $\Fex (\D)$ and its approximation $\Fapx (\D)$,  for three different choices of two-qubit gates, (a),(d) $\CZ$; (b),(e) $\fSim (\pi/2, \pi/6)$; and (c),(f) 2HR.
    (This correspondence between a column of panels and a choice of two-qubit gates will be also used in later figures.)
      Panels (a)--(c) [panels (d)--(f)] show the fidelity per qubit $\mc{F}^{1/n}$ [the effective error per two-qubit gate $\ve = 1 - n_\mr{2qg}^{-1} \log \mc{F}$], where $n_\mr{2qg}(\D) \approx (2n - n_r - n_c) (\D /4)$ means the number of two-qubit gates applied up to depth $\D$.
      The average and the standard deviation of the data obtained for ten different circuit instances are depicted as dots (solid lines) and error bars (shades) for $\Fex$ ($\Fapx$), respectively.
      Line colors distinguish different values of the maximum bond dimension $\chi$.
      In panels (d)--(f), black dashed lines represent $\ve = 1 - 2^{-n/n_\mr{2qg}} \approx \D^{-1} \log 4$, associated with the fidelity saturation $\mc{F} = 2^{-n}$.
      Here $\varepsilon$ derived from $\Fapx$ is not shown beyond the saturation line since the PEPS representation is no better than a random guess in that regime.
    }
    \label{fig:exact_fid}
    \vspace{-1em}
\end{figure*}

The three-stage dependence is further supported by \Fig{fig:exact_fid}, which shows $\Fex(\D)$ for other choices of two-qubit gates and $\chi$'s.
We find that $\Dtr \sim \log \chi$ increases with $\chi$, while $\elayer$ is independent of $\chi$.
While the overall behaviors of $\Fex (\D)$ are similar, there are small but noticeable deviations between different gate sequences.
The curves for $\CZ$ and $\fSimpp$ show sharp transition from $\Fex = 1$ to the exponential decay with small steplike wiggles.
On the other hand, the curves for 2HR gates show smooth crossovers to decay with almost no wiggles.
We explain this gate dependence in \Sec{sec:Fex vs Fapx}, where we discuss the behavior of $\Fapx (\D)$.

\subsection{nXEB}
\label{sec:nXEB}

Before moving on to the discussion on the approximate fidelity $\Fapx$, we examine the relation between $\Fex$ and the nXEB~\cite{Dalzell2024, Ware2023}, which is defined by
\begin{equation}
\Fnxeb = \frac{2^n \sum_{\mb{x} \in \{ 0, 1\}^n} p_\peps (\mb{x}) p_\mr{ex} (\mb{x}) - 1}{2^n \sum_{\mb{x} \in \{ 0, 1\}^n} p_\mr{ex}^2 (\mb{x}) - 1} ,
\label{eq:nXEB}
\end{equation}
where $p_\mr{ex} (\mb{x}) = | \ovl{\mb{x}}{\psi_\mr{ex}} |^2$ and $p_\peps (\mb{x}) = | \ovl{\mb{x}}{\psi_\peps} |^2$ are the bitstring probabilities according to, respectively, the exact state $\ket{\psi_\mr{ex}}$ and the PEPS $\ket{\psi_\peps}$.
The numerator is the linear XEB~\cite{Arute2019short}, which can be much larger than 1 when $\D$ is small.
The denominator normalizes the overshooting for small $\D$, while it becomes $1$ for large $\D$ where the distribution of $p_\mr{ex}$ follows the Porter--Thomas distribution (PTD) $\Prob (p) = 2^n \mr{exp} (-2^n p)$.
The XEB family is of particular interest in RQC sampling experiments for two reasons: 
First, it is regarded as a proxy for fidelity, which can be estimated via sampling~\cite{Boixo2018, Arute2019short}.
Second, achieving its high value is considered a quantum advantage task~\cite{Aaronson2020}.
However, Refs.~\cite{Ayral2023,Gao2024} criticized the first aspect, showing that the XEB can be much larger than the fidelity.

Figure~\ref{fig:XEB vs fidelity} reveals that the relation between $\Fex$ and $\Fnxeb$ in PEPS calculations depends on $\chi$.
When $\chi$ is small, $\Fnxeb$ can be much larger than $\Fex$.
We find that $\Prob (p_\peps)$ differs from the PTD in this case (\Fig{fig:PTD}), so the assumption underlying the use of the nXEB as a fidelity proxy does not hold.
As $\chi$ increases, $\Prob (p_\peps)$ converges to the PTD (\Fig{fig:PTD}). Accordingly, the curve of $\Fnxeb$ versus $\Fex$ approaches two-stage dependence,
\begin{equation}
\Fnxeb \simeq
\begin{cases}
\Fex, & \Fex > 2^{-n/2}, \\
2^{-n/2}, & \text{otherwise},
\end{cases}
\end{equation}
which means that $\Fnxeb$ is a faithful proxy of $\Fex$ when $\chi$ is large enough and $\Fnxeb$ is larger than its saturation limit $2^{-n/2}$.
The $\D$ dependence of $\Fnxeb$ for such large $\chi$ is plotted in Figs.~\ref{fig:angles}(a) and~\ref{fig:angles}(b).
The saturation limit of $\Fnxeb = 2^{-n/2}$, which is the square root of the saturated fidelity $\Fex = 2^{-n}$, is also observed in MPS numerics and justified based on random matrix theory~\cite{Ayral2023}.

\subsection{Approximate fidelity}
\label{sec:Fex vs Fapx}

As discussed in \Sec{sec:Fapx}, one of the major advantages of tensor network approaches is that one can calculate the approximate fidelity $\Fapx$ without computing the exact state~\cite{Ayral2023, Zhou2020, Sun2024, Shaw2024}.
Figures~\ref{fig:exact_fid}(a)--\ref{fig:exact_fid}(c) reveal that $\Fex$ and $\Fapx$ agree well when $\D < \Dsat$.
While $\Fex$ saturates to $2^{-n}$ for $\D > \Dsat$, $\Fapx$ keeps decaying since the bonds are consistently truncated.
Thus, $\Fapx$ exhibits the two-stage dependence on $\D$,
\begin{equation}
\Fapx (\mc{D}) \simeq
\begin{cases}
1, & \mc{D} < \Dtr, \\
\me^{- \elayer (\mc{D} - \Dtr)}, & \Dtr \leq \mc{D} .
\end{cases}
\label{eq:Fapx_D}
\end{equation}
The agreement validates the use of $\Fapx$ for large-scale representation in which $\Fex$ is not available (cf.~\Sec{sec:scalability}).

Since $\Fapx (\D)$ is a good approximation of $\Fex$ and it depends only on the singular values $\{ \sigma_{\underline{1}}^{(i,j)} \}$, the gate-dependent behaviors of $\Fex (\D)$ mentioned in the previous subsection can be explained accordingly.
The singular values are highly degenerate for $\fSimpp$, less degenerate for $\CZ$, and non-degenerate for 2HR gates [see \Fig{fig:degeneracy} and \hyperref[app:entanglement]{Appendix}.
(The singular value degeneracy in a PEPS generated by the $\CZ$ gate sequence is also reported in Ref.~\cite{Guo2019}.)
In the $\CZ$ or $\fSimpp$ case, the discarded weight as a function of $\chi$ is discrete, as the singular value spectrum is highly discrete;
the sharp drop of the fidelity at $\D = \Dtr$ and the steplike wiggles in the decaying regime reflect the discreteness.
On the other hand, the singular value spectrum in the 2HR case is smooth, so $\Fex(\D)$ and $\Fapx(\D)$ are smooth.

The singular value degeneracy is directly linked to the degeneracy of the operator Schmidt coefficients (OSCs) of each two-qubit gate.
Here, the OSCs of a gate $g = \sum_{x_1, x_2, x'_1, x'_2} g^{x_1 x_2}_{x'_1 x'_2} \ket{x_1 x_2}\!\bra{x'_1 x'_2}$ are defined as the Schmidt coefficients of its Choi-Jamio{\l}kowski isomorphism $\ket{g} =  \sum_{x_1, x_2, x'_1, x'_2} g^{x_1 x_2}_{x'_1 x'_2} \ket{x_1 x'_1} \otimes \ket{x_2 x'_2}$.
The OSCs for $\CZ$ and $\fSim$ are $\{ 1, 1 \}$ and $\{ 1, 1, 1, 1 \}$, respectively, while they are nondegenerate for 2HR in general.

\begin{figure*}
    \centering{\includegraphics[width=\textwidth]{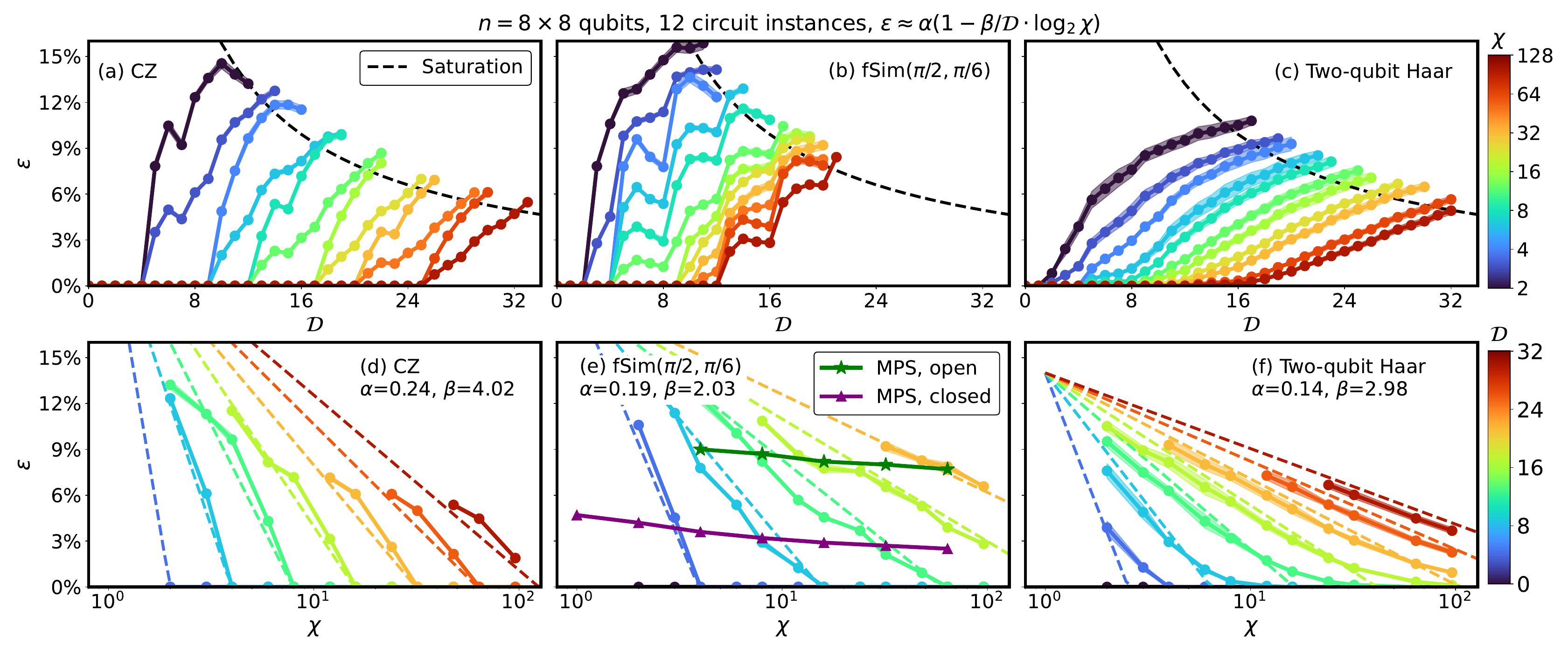}}
    \caption{
    (a)--(c) Error per two-qubit gate $\ve$ derived from $\Fapx$, plotted as a function of $\D$ for various values of $\chi$.
    (d)--(f) The same data (for $\D$'s that are multiples of $4$) is plotted differently, as a function of $\chi$ for various values of $\D$.
    In all panels, solid lines (accompanying shades) show the average (standard deviation) of the results obtained for 12 circuit instances.
    Black dashed lines in panels (a)--(c) indicate $\ve \approx \D^{-1} \log 4$ associated with the saturation limit $\Fapx = 2^{-n}$.
    Colored dashed lines in panels (d)--(f) depict the scaling relation $\ve = \alpha [1 - (\beta / \D) \log_2 \chi ]$ with $\alpha$ and $\beta$ estimated by fitting the PEPS results of $\ve$ to the relation.
    In panel (e), the green line with stars (purple line with triangles) represents the open (closed) MPS results taken from Fig.~2(a) of Ref.~\cite{Ayral2023}.
    Note that the MPS simulations were for $(n, \D) = (54, 20)$ on the Sycamore-like rotated lattice, which is computationally easier than $n = 8 \times 8$ qubits on the standard square lattice we consider here.
    }
    \label{fig:8x8 fidelity}
    \vspace{-1em}
\end{figure*}

To compare with other theoretical and experimental results, we convert the fidelity $\mc{F}$ into an effective error per two-qubit gate $\ve = 1 - n_\mr{2qg}^{-1} \log \mc{F}$ (where $n_\mr{2qg}$ denotes the number of two qubit gates used; cf.~\Sec{sec:random_quantum_circuit}), based on the assumption that the single-qubit gates are noiseless and the two-qubit gates have uncorrelated noises $\ve$, leading to the fidelity decrease $\mc{F} = (1 - \ve)^{n_\mr{2qg}} \approx \mr{e}^{-\ve (2n - n_r - n_c) \D / 4}$.
While $\Fex$ and $\Fapx$ decay exponentially at $\D > \Dtr$, the corresponding increase of $\ve (\D > \Dtr)$ is sublinear.

\section{Error scaling in large-scale representation}
\label{sec:scalability}

We demonstrate a computational advantage of the PEPS approach in representing RQC states for large $n$.
We evaluate its performance based on the error per two-qubit gate $\ve$ derived from $\Fapx$, as the computation of $\Fex$ is hard for large $n$ or large $\D$ (cf.~\Sec{sec:Fapx}).

In Figs.~\ref{fig:8x8 fidelity}(a)--\ref{fig:8x8 fidelity}(c), we show the results for $n = 8 \times 8$ qubits, where the state-vector simulation is unavailable.
For a given $\chi$, $\ve (\D)$ stays at $0$ until $\D$ reaches $\Dtr$, then it increases before crossing the fidelity saturation line $\ve\approx\D^{-1} \log 4$.
When one chooses a larger $\chi$, the truncation depth $\Dtr$ increases, and the overall slope after $\Dtr$ lowers.
Note that the behavior of $\ve (\D \geq \Dtr)$ can be nonmonotonous as shown in Figs.~\ref{fig:8x8 fidelity}(a) and~\ref{fig:8x8 fidelity}(b) for the $\CZ$ and $\fSimpp$ sequences, which reflects the steplike wiggles of $\mc{F} (\D)$ originating from the singular value degeneracy (cf.~\Sec{sec:Fex vs Fapx}).
To rule out such wiggles in the scaling analysis, we plot the data for $\D \equiv 0 \, (\mr{mod} \, 4)$ as a function of $\chi$, in Figs.~\ref{fig:8x8 fidelity}(d)--\ref{fig:8x8 fidelity}(f).
These plots show the following scaling behavior:
\begin{equation}
\ve (\chi) \simeq \max \left[
\alpha\left(1-\frac{\beta}{\mc{D}}\log_2\chi\right), 0
\right],
\label{eq:error per gate}
\end{equation}
where $\alpha$ and $\beta$ depend on the choice of two-qubit gates.
These are related to $\Dtr$ and $\elayer$ as
\begin{equation}
\Dtr = \beta \log_2 \chi, \quad \elayer = \alpha n /2 .
\end{equation}
That is, the error scaling of \Eq{eq:error per gate} is consistent with the fact that bond dimensions of an untruncated PEPS grow exponentially as $\D$ increases.
We estimate $(\alpha, \beta) \simeq (0.24, 4.02),\,(0.19, 2.03)$, and $(0.14, 2.98)$, for the $\CZ$, $\fSimpp$, and 2HR gates, respectively, by fitting the $n = 8 \times 8$ data to \Eq{eq:error per gate}.
As the value of $\chi$ giving $\ve = 0$ is solely determined by $\beta$ in \Eq{eq:error per gate}, we can rank the computational difficulties of the three gate sequences based on the values of $\beta$:
The $\fSimpp$ case is the hardest and the $\CZ$ case is the easiest.
This ranking can also be confirmed by comparing the curves from Figs.~\ref{fig:8x8 fidelity}(a)--\ref{fig:8x8 fidelity}(c) for a fixed $\chi$.

\begin{figure*}
    \centering
    \includegraphics[width=\textwidth]{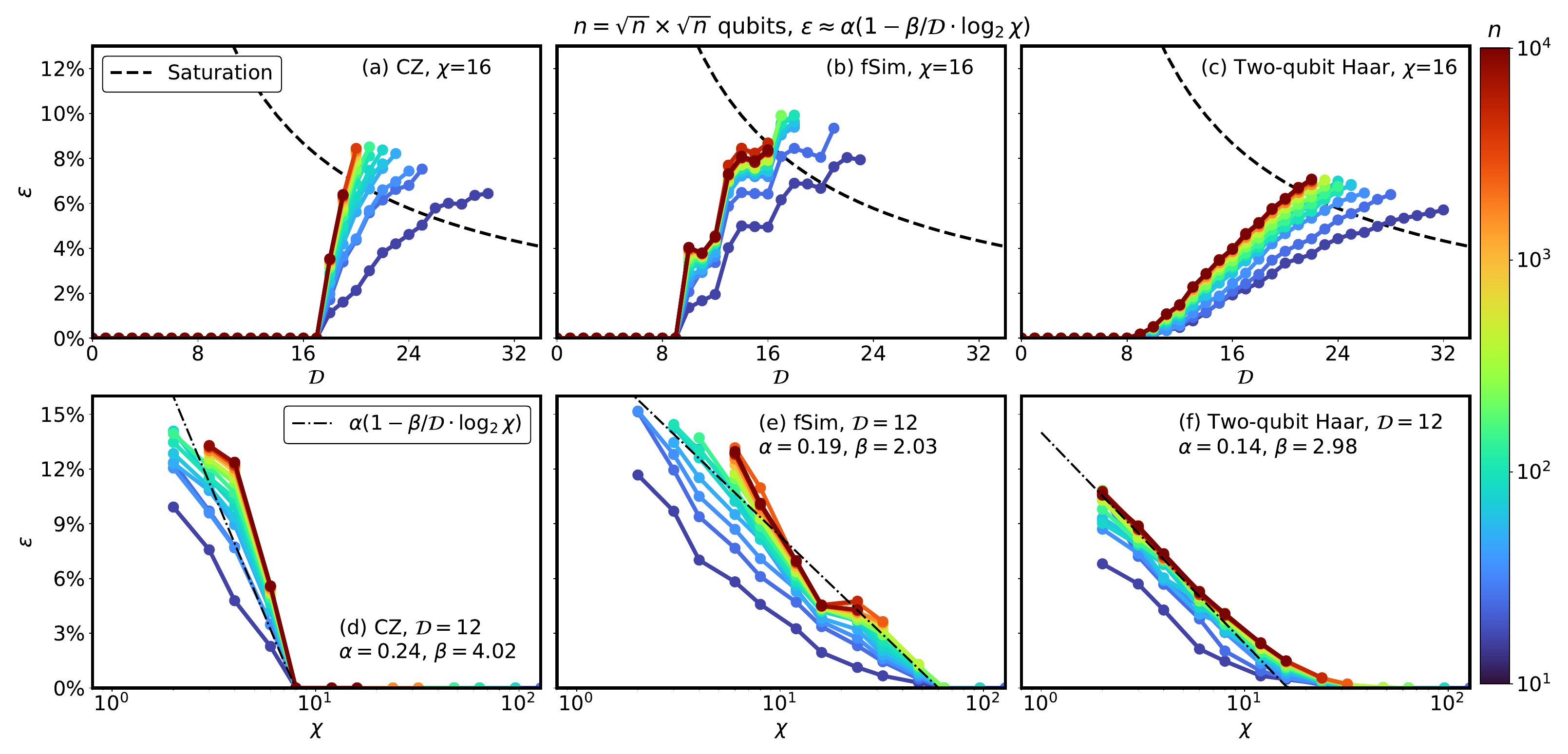}
    \caption{
    (a)--(c) Error per two-qubit gate $\varepsilon$ as a function of $\D$, for different number of qubits $n$ and a fixed $\chi$.
    (d)--(f) $\varepsilon$ as a function of $\chi$, for different number of qubits $n$ and a fixed $\D$. In all panels, solid lines with dots show the average saturation limit of the results obtained for ten circuit instances.
    In panels (d)--(f), dash-dotted lines indicate the error scaling relation $\ve = \alpha [1-(\beta/\mc{D}) \log_2\chi ]$ for the parameters $(\alpha, \beta)$ estimated by fitting the results for $n = 8 \times 8$ qubits in \Fig{fig:8x8 fidelity}.
    }
    \label{fig:universality_by_bond}
    \vspace{-1em}
\end{figure*}

In \Fig{fig:universality_by_bond}, we further extend the analysis to square-sized qubit lattices with $n = \sqrt{n} \times \sqrt{n}$ for a wide range of $16 \leq n \leq 10^4$.
The behavior $\ve (\chi)$ in Figs.~\ref{fig:universality_by_bond}(d)--\ref{fig:universality_by_bond}(f) follows the scaling in  \Eq{eq:error per gate}.
We find that $\beta$, which determines $\chi$ such that $\ve(\chi) = 0$, is almost independent of $n$, while $\alpha$, which determines the slope of $\ve$ versus $\log\chi$, has a weak dependence on $n$.
The $\ve (\chi)$ curves for $n \gtrsim 10^3$ lie on top of each other, which reveals the universality in the error scaling in the large-$n$ limit.

The $\log \chi$ behavior of $\ve(\chi)$ in \Eq{eq:error per gate} also appears in the MPS-based simulations of RQC states~\cite{Zhou2020,Ayral2023,DeCross2024short}.
In Ref.~\cite{Ayral2023}, the DMRG calculation for the Sycamore-like circuits manifests the error scaling at the chaotic limit,
\begin{align}
&\ve_\mps \simeq \frac{1}{\D} \left( \log 2 - \frac{\log 4\chi}{n/2} \right),
\label{eq:error_mps_Ayral}
\end{align}
which can be derived analytically by assuming that the coefficients of the $n$-qubit state vector are given by the elements of a $2^{n/2} \times 2^{n/2}$ random Gaussian matrix.

\begin{table}
\renewcommand{\arraystretch}{1.5}
\caption{Performance of our PEPS method for representing an $n$-qubit RQC state at depth $\D$.
Here, $z$ is the coordination number of the qubit lattice, and the gate-dependent coefficients $\alpha$ and $\beta$ are obtained via regression (cf.~\Fig{fig:8x8 fidelity}).
The third and fourth rows indicate the scaling behaviors, not strict relations.}
\begin{tabular*}{\linewidth}{@{\extracolsep{\fill}} l l}
\hline\hline
Time complexity & $O\big(\mc{D} n \chi^{z+1}\big)$ \\
Space complexity & $O\big( n\chi^z \big)$ \\
$\chi$ for exact PEPS representation ($\ve = 0$) & $\simeq 2^{\mc{D}/\beta}$ \\
Error per two-qubit gate $\ve$ & $\simeq \alpha\left(1-\frac{\beta}{\mc{D}}\log_2\chi\right)$ \\
\hline\hline
\end{tabular*}
\label{tab:performance}
\end{table}

Based on the $\log \chi$ behaviors that appear in both MPS and PEPS calculations, we can benchmark their computational cost required for error-free representation, i.e., $\ve = 0$.
For this, we compare our PEPS result for $(n, \D) = (64, 20)$ (orange line) and the MPS result for $(n, \D) = (54, 20)$~\cite{Ayral2023} (purple line), both plotted as a function of $\chi$ in \Fig{fig:8x8 fidelity}(e).
When the $\ve$ versus $\log \chi$ dependencies are linearly extrapolated, $\ve$ becomes $0$ for $\chi  \simeq 10^3$ and $\simeq 10^5$ for the PEPS and MPS cases, respectively.
By substituting the values into the time complexity of our PEPS method given in Table~\ref{tab:performance}, one gets $O(10^{16})$ floating point operations.
In contrast, the time complexity of the DMRG approach used in Ref.~\cite{Ayral2023} scales as $O\big(\D L \chi^2 2^{n/L} \big)$, where $L = 3$ is the number of MPS tensors, each of which is associated with a subset of qubits.
Substituting the numbers, its time complexity is on the order of $O (10^{17})$.
Given that we ignore prefactors and subleading contributions, we can say that 
our PEPS method is not much slower (unless faster) than the MPS approaches in obtaining the exact state representation.
This advantage comes from that the PEPS inherits the geometry from the lattice to solve;
its generalization to random geometry~\cite{DeCross2024short} will be an interesting topic for future research.

For even larger systems in two spatial dimensions, the PEPS approach would become more favorable since the computational cost for the MPS method increases exponentially with the shorter length (see the factor $2^{n/L}$ in the time complexity), while the PEPS method's time and space complexities are linear in $n$.
Furthermore, the simple update algorithm concerns tensors only locally, so it can be highly parallelized.

\vspace{2.5em}

\section{Conclusion and Outlook}
\label{sec:conclusion}

We represented the RQC states as PEPSs in the Vidal gauge, treating their evolution along the circuits with the simple update method.
As an ansatz that directly reflects the qubit lattice geometry, the PEPS representation can encode the entanglement generated by nearest-neighbor two-qubit gates by manipulating tensors only locally, hence scalable in terms of the number of qubits $n$ up to logarithmic depth $\D=c\log{n}$, regardless of the choice of single- and two-qubit gates.
The Vidal-gauge conditions provide an efficient way to estimate the state fidelity.
The fidelity has a systematic dependence on the maximum bond dimension $\chi$ and the choice of two-qubit gates,
and the effective error per two-qubit gate derived from the fidelity manifests the universal scaling in the limit of large $n$.
While $\chi$ needed for error-free exact representation increases exponentially with the circuit depth, we could handle $n = 8 \times 8$ up to $12 \lesssim \D \lesssim 24$
(depending on two-qubit gates) by using a conventional CPU for O(1) core-hours per circuit instance.
The PEPS method has lower computational costs than the MPS approaches, especially when $n \gtrsim 50$.
Note that another computational advantage of the Vidal-gauge PEPS has recently been demonstrated~\cite{Tindall2024,Begusic2024,Patra2024} in benchmarking IBM's kicked Ising experiment~\cite{Kim2023}.

Below we discuss the performance of the PEPS method in the context of entanglement scaling laws (\Sec{sec:entanglement_laws}), physical observables (\Sec{sec:local_and_global}), and how far it can reach, if it is leveraged with high-performance computing techniques (\Sec{sec:log_depth}).

\subsection{Entanglement scaling laws}
\label{sec:entanglement_laws}

An RQC with sufficiently large $\D$ is expected to draw a typical state from the Haar-random unitaries, which has a volume-law entanglement, i.e., a bipartite entanglement with magnitude proportional to the number of qubits in a smaller subsystem.
Indeed, we confirm that the RQC states have volume-law entanglement in later depths, from state-vector simulations (cf.~\Fig{fig:entanglement entropy} and \hyperref[app:entanglement]{Appendix}).
On the other hand, the PEPS with constant $\chi$ can capture only up to an area-law entanglement whose magnitude is proportional to the number of bonds sitting between the two subsystems.
So our claim of the scalable PEPS representation may sound contradictory to the volume-law entanglement of a Haar random state.

What is important here is that an RQC should be viewed as a dynamic system, whose initial state has zero entanglement and thus has sub-volume-law entanglement for smaller $\D$.
Accordingly, the PEPS can exactly describe an RQC state up to a finite depth, until the entanglement grows beyond its expressivity set by $\chi$.
Once the expressivity limit is reached, truncations happen and the fidelity decays exponentially.
Though the exponential decay seems daunting, the equivalent error per two-qubit gate increases only linearly as a function of $\D$, which makes the PEPS calculation competitive against noisy experiments.

\begin{figure*}
    \centering
    \includegraphics[width=\textwidth]{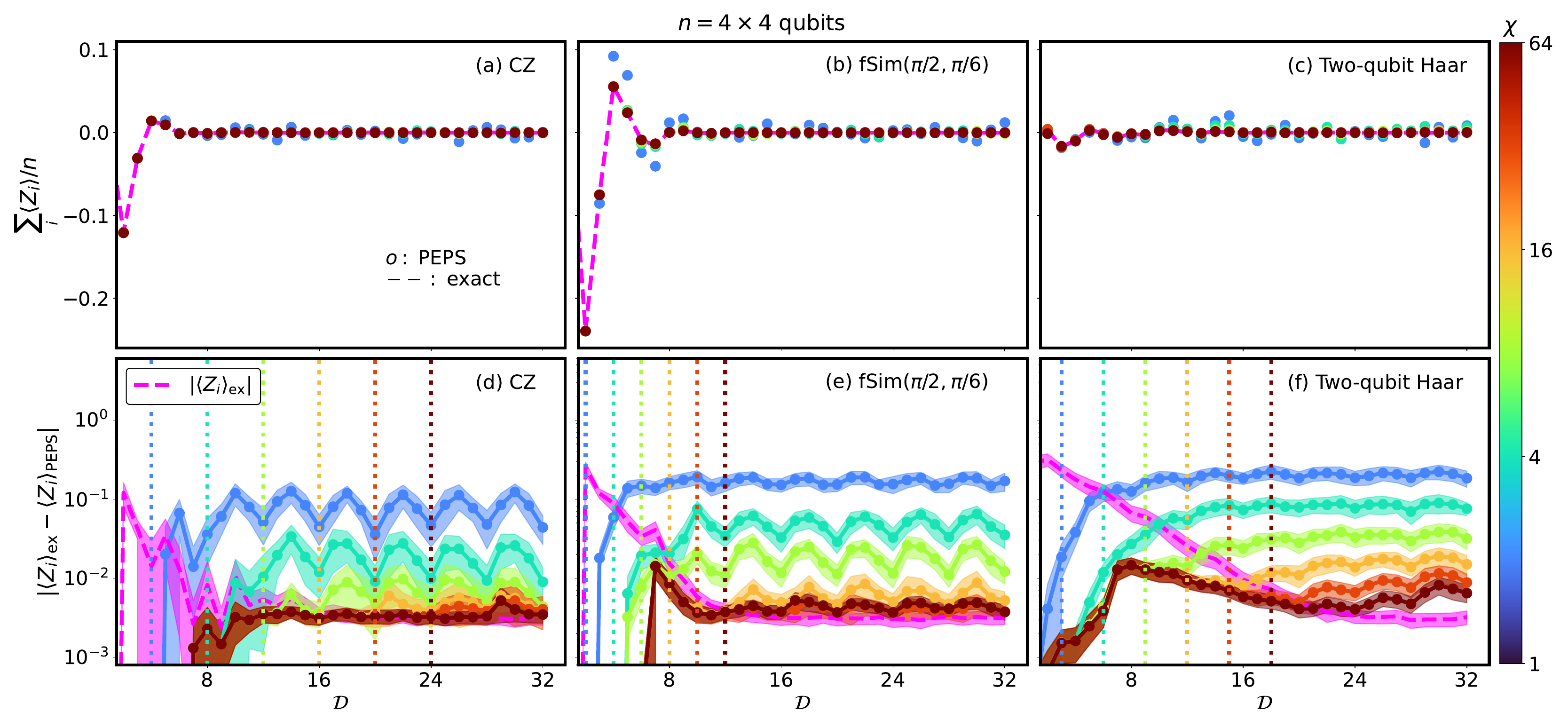}
    \caption{
    (a)--(c) The expectation values of the local Pauli-$Z$ measurements, $\la Z_i \ra$, averaged over $n = 4 \times 4$ qubits, plotted as a function of $\D$.
    Dashed lines show the exact values obtained by state-vector simulations.
    Dots are from the approximate local contraction based on the Vidal gauge, depicted in \Fig{fig:PEPS}(e), while their colors distinguish $\chi$.
    (d)--(f) Solid lines with dots represent the absolute error of the PEPS results, $| \la Z_i \ra_\peps - \la Z_i \ra_\mr{ex} |$, averaged over qubits, where colors code $\chi$.
    Magenta dashed lines represent the average of $| \la Z_i \ra_\mr{ex} |$; the error lower than $| \la Z_i \ra_\mr{ex} |$ implies that the PEPS is better than a naive guess $\la Z_i \ra = 0$.
    Shades accompanying these lines indicate the corresponding standard deviations.
    }
    \label{fig:Pauli_Z}
\end{figure*}

\subsection{Physical observables}
\label{sec:local_and_global}

A quantum state cannot be directly determined in experiments, unless one performs quantum state tomography, whose cost increases exponentially with the number of qubits.
Therefore, a classical simulation is benchmarked with quantum hardware by measuring observables.
In tensor network methods, the expectation value of an observable is obtained by contracting tensor networks.
While such contraction is straightforward for MPSs, it is generally nontrivial for PEPSs because of loops therein.
However, with proper approximation of the environment, one can efficiently extract the expectation value of local observables from a PEPS with high accuracy, as showcased in many applications to strongly correlated systems (see, e.g., Ref.~\cite{Zheng2017}).

As an example relevant to digital quantum circuits, we compute the expectation value of local Pauli-$Z$ measurement $\la Z_i \ra$, approximating the environment outside of site $i$ to the $\Lambda$ tensors, as described in \Fig{fig:PEPS}(e).
Figure~\ref{fig:Pauli_Z} shows that the absolute error $| \la Z_i \ra_\mr{ex} - \la Z_i \ra_\peps |$ of this simple calculation (with computational complexity comparable to each simple update step) can be systematically decreased by increasing $\chi$.
The error can be finite even when the PEPS has fidelity $\mc{F} = 1$ for $\D < \Dtr$, since the correlated environment was approximated to an uncorrelated one.
Accordingly, the accuracy can be further improved by incorporating a larger correlated environment, as in the cluster and full updates~\cite{Lubasch2014, Lubasch2014a}, or by using the loop-corrected belief propagation method~\cite{Evenbly2024,Tindall2025}.

Nonlocal observables can be decomposed into the sum of weight-$m$ observables, i.e., the product of Pauli operators acting on $m$ different qubits.
Each weight-$m$ observable can be transformed into a single $Z$ via Clifford gates.
That is, by applying the inverse of the Clifford gates to a PEPS, then by performing a local $Z$ measurement with the updated PEPS, one obtains the expectation value of the weight-$m$ observable.
Employed in Refs.~\cite{Tindall2024,Patra2024}, this approach is shown to outperform IBM's quantum hardware.

A global observable, which is of interest in the context of RQC sampling, is the bitstring probability, $p(\mb{x}) = |\ovl{\mb{x}}{\psi} |^2$, since the strong simulation, i.e., computing $\{ p(\mb{x}) \}$ with small total variation distance $\epsilon_{\scriptscriptstyle{\mr{TVD}}} = (1/2) \sum_{\mb{x} \in \{ 0, 1\}^n} | p_\mr{ex} (\mb{x}) - p_\peps (\mb{x})|$, is considered to be asymptotically hard for RQCs~\cite{Hangleiter2023}.
A conventional PEPS-based approach for $p(\mb{x})$ is to project the physical legs onto $\{ \ket{x_i } \}$ and then to contract the resulting two-dimensional tensor network using the boundary MPS.
However, when the PEPS has volume-law entanglement (cf.~\Sec{sec:entanglement_laws}), this approach fails because of exponentially large bond dimensions of the boundary MPS.
We expect the loop-corrected belief propagation will be beneficial in overcoming this issue, as shown in a recent benchmark against the boundary MPS method about the quantum annealing dynamics of Ising spin glasses~\cite{Tindall2025}.

\subsection{Computational costs for log-depth circuits}
\label{sec:log_depth}

\begin{figure*}
    \centering
    \includegraphics[width=\textwidth]{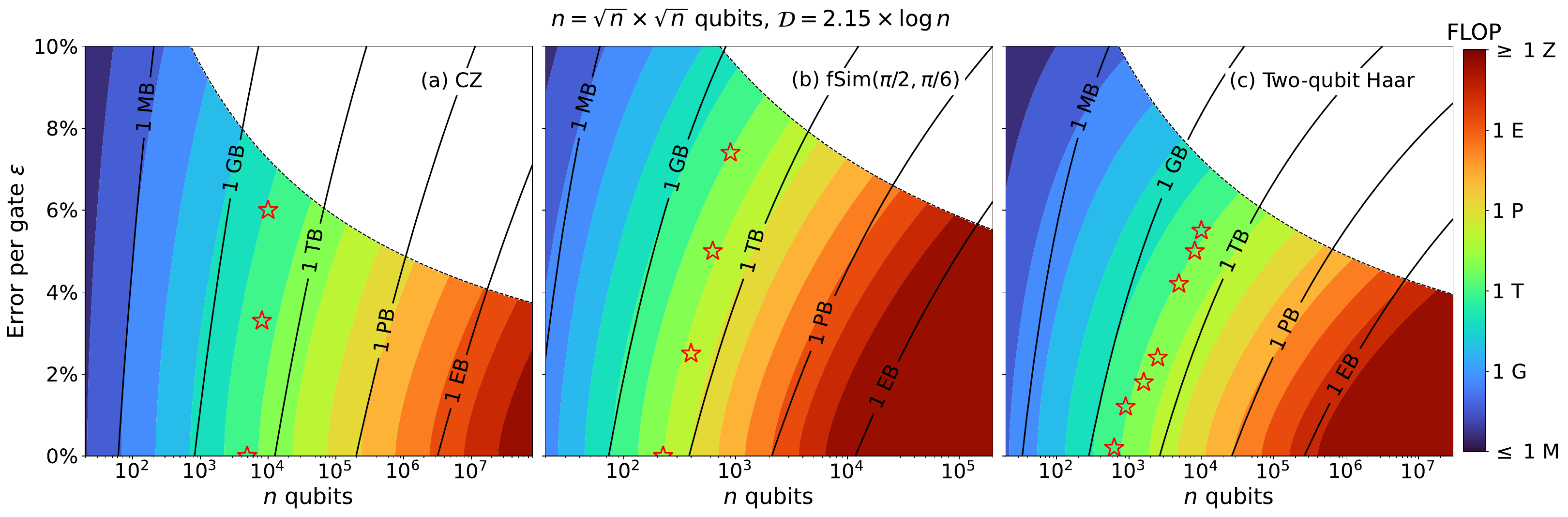}
    \caption{
    Estimated computational resources for representing an RQC state at depth $\mc{D}= 2.15 \times \log n$, as a function of the number of qubits $n$ and the desired error per two-qubit gate $\ve$.
    Color gradients indicate floating point operations in the log scale.
    Black solid lines are the contours of memory requirements on a log scale.
    Red stars represent the heaviest computational costs performed in this study, which are mostly limited by the 128 GB memory of the CPU machine we used.
    In white regions on the upper right corners, state fidelities are smaller than the random guess limit $2^{-n}$, so any simulation becomes futile.
    }
    \label{fig:simulable_regime_log}
\end{figure*}

RQCs of depth $\D = c \log n$ with $c = O(1)$, called log-depth circuits, have been of particular interest in the theory of RQC in the context of \textit{anticoncentration} of bitstring probabilities $p(\mb{x)}$.
The probabilities, which were concentrated to a single bitstring $p(\mb{x}) = \delta_{\mb{x},\mb{0}}$ in the initial state $\ket{\mb{0}}$, spreads over a large subset of bitstrings~\cite{Aaronson2013} through a log-depth circuit~\cite{Dalzell2022, Barak2021}. 
This anticoncentration underlies complex-theoretic proofs and arguments that quantifying bitstring probabilities is hard using classical algorithms.
For 2HR gates, the value of $c$ is found to be $5/3 \simeq 1.67$ and $1/\log(5/4) \simeq 4.48$ for a complete graph and a one-dimensional geometry, respectively~\cite{Dalzell2022, Barak2021}.

We find that our observation of the 2HR case is consistent with the anticoncentration at a depth $\D = c \log n$ with $c \simeq 2.15$.
A random PEPS, whose constituent tensors are randomly drawn from the Haar ensemble for a given bond dimension $\chi$, exhibits anticoncentration when $\chi \gtrsim \sqrt{n}$~\cite{Lami2025}.
On the other hand, in our study, a PEPS constructed using the simple update method has bond dimensions $\chi \simeq 2^{\D/\beta}$, when bonds are not truncated.
If the latter---a PEPS obtained via the simple update---can be viewed as a random PEPS, then anticoncentration occurs at depth $\D_\mr{ac}$ such that $2^{\D_\mr{ac} / \beta} \simeq \sqrt{n}$.
(In the case of one-dimensional circuits, the similarity between the MPS representations of RQC states and random MPSs is investigated in a recent work~\cite{Sauliere2025}.)
Accordingly, the anticoncentration depth is given by $\D_\mr{ac} \simeq \beta/(2 \log 2) \log n$, which leads to the $c \log n$ depth with $c \simeq 2.15$ as $\beta \simeq 2.98$ for 2HR gates.

For log-depth circuits, the bond dimension for error-free representation is linear in $n$, so the time and space complexities of our PEPS method scale as $O(n^{(z+1)c\log{2}/\beta+1} \log n)$ and   $O(n^{zc\log{2}/\beta+1})$, respectively (cf.~Table~\ref{tab:performance}).
The polynomial cost scaling means that the PEPS method is efficient and can be further scaled up for much larger $n$ if one has access to computational resources, as we estimated in \Fig{fig:simulable_regime_log}.
We find that the easiness for log-depth circuits is consistent with the result of a recent complex-theoretic study~\cite{Deshpande2022}.

\begin{acknowledgments}
We thank Philippe Corboz, Gyungmin Cho, Byeongseon Go, Chu Guo, Mingyu Kang, Fabian Kugler, Wonjun Lee, Changhun Oh, Tsuyoshi Okubo, Seongwook Shin, Feng-Feng Song, Jan von Delft, Xavier Waintal, and Pan Zhang for fruitful discussions.
This work was supported by the National Research Foundation of Korea (NRF) grants funded by the Korean government (MSIT) (No.~RS-2023-00214464, No.~RS-2023-00258359, No.~RS-2023-NR119931, No.~RS-2024-00442710), the Global-LAMP Program of the NRF grant funded by the Ministry of Education (No.~RS-2023-00301976), the NRF grant funded by the Korean government (MEST) (No.~2019R1A6A1A10073437), and Samsung Electronics Co., Ltd.~(No.~IO220817-02066-01).
S.-B.B.L. was supported by the Korea Institute for Advancement of Technology (KIAT) grant funded by the Korean Government (Ministry of Education) (P0025681-G02P22450002201-10054408, Semiconductor-Specialized University).
S.-B.B.L. and D.D.O. were also supported by the SNU Student-Directed Education Undergraduate Research Program through the Faculty of Liberal Education, Seoul National University (2024).

\end{acknowledgments}

\begin{figure*}
    \centering
    \includegraphics[width=\textwidth]{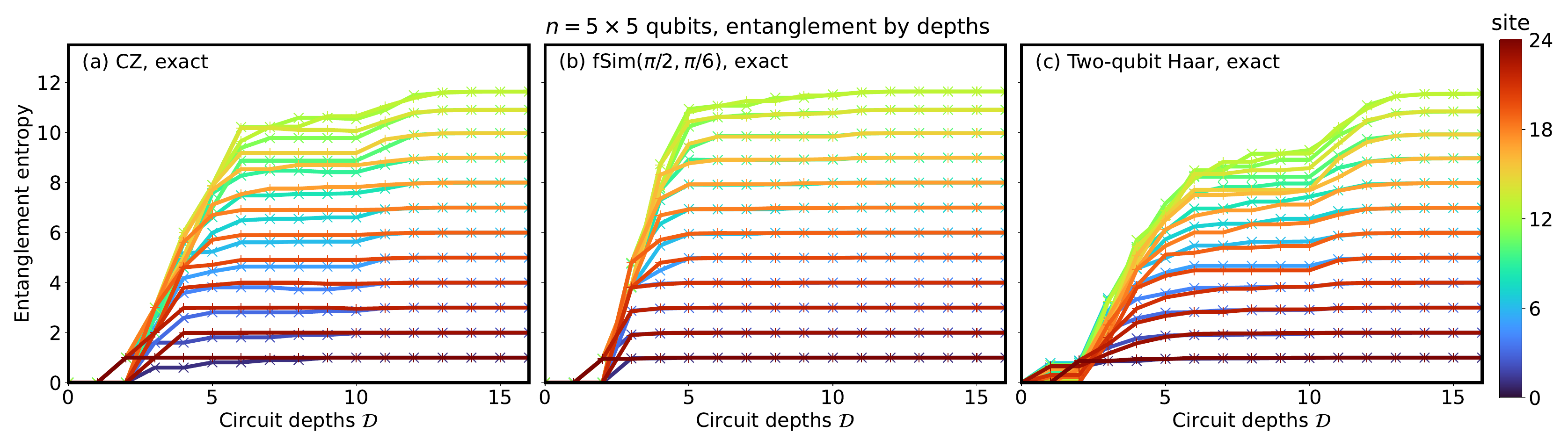}
    \caption{
    Entanglement entropy $S_i = - \tr ( \rho_{1, \dots, i} \log_2 \rho_{1, \dots, i} )$ between one subsystem (sites $1, \dots, i$) and the rest (sites $i+1,\dots, n$) as a function of $\D$, obtained from exact state-vector simulations.
    Different line colors encode the qubit site index $i$, where the sites are indexed in the English reading order in the $n = 5 \times 5$ lattice.
    Here we show the result of one circuit instance for each choice of two-qubit gates.
    }
    \label{fig:entanglement entropy}
\end{figure*}

\begin{figure*}
    \centering
    \includegraphics[width=\textwidth]{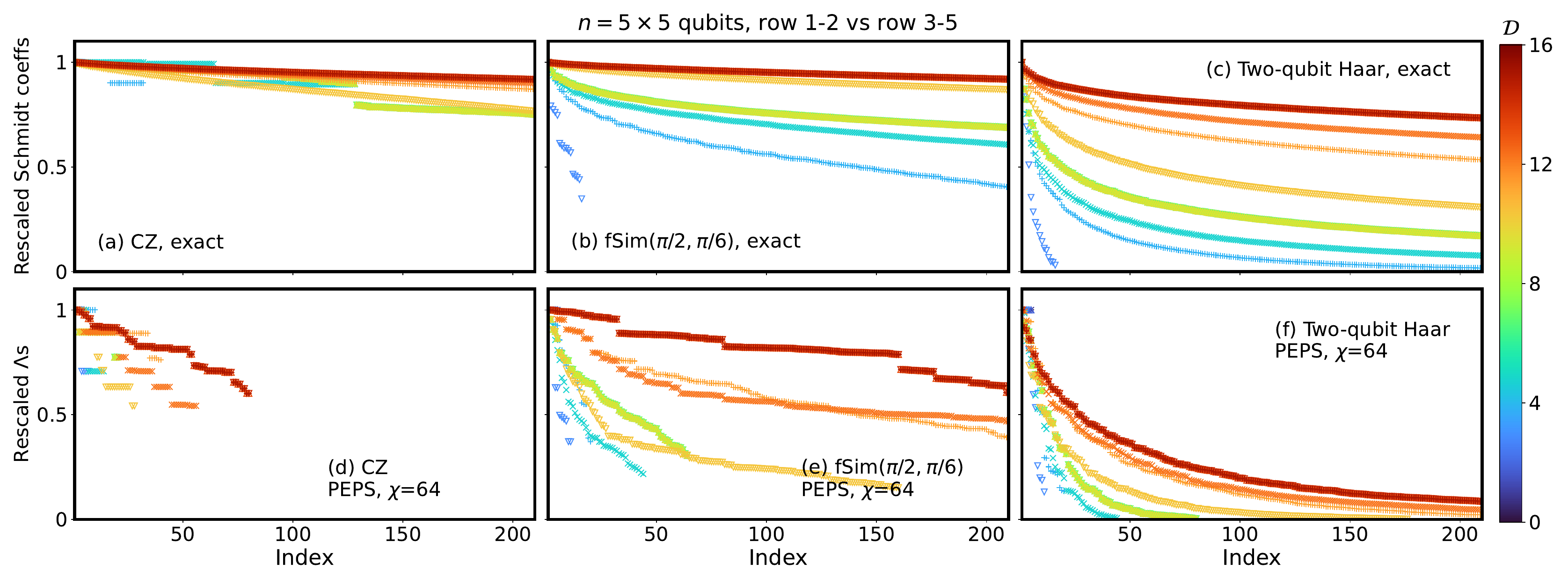}
    \caption{
    (a)--(c) Entanglement spectra (i.e., Schmidt coefficients) associated with the entanglement entropy between rows 1--2 and rows 3--5, in state-vector simulations.
    (d)--(f) Diagonal elements of the $\Lambda$ tensors lying between row 2 and row 3, in PEPS calculations.
   Different line colors indicate different depth $\D$.
   Both entanglement spectrum and $\Lambda$ tensor elements are sorted in nonincreasing order and rescaled by dividing by the largest element so that the first data point is $1$.
   We perform exact PEPS representation for depths $\D \leq 16$ shown here, by choosing $\chi = 64$.
    }
    \label{fig:degeneracy}
\end{figure*}

\appendix*
\section{Entanglement properties of RQC states}
\label{app:entanglement}

Since the performance of tensor network approaches is intimately related to entanglement structure, it is necessary to understand the entanglement properties of RQC states.
In \Fig{fig:entanglement entropy}, we plot how the bipartite entanglement entropy $S_i$ evolves with the circuit depth for $n = 5 \times 5$ qubits.
For the $\CZ$ and 2HR sequences, the entanglement entropy grows linearly for $\D \lesssim 6$ and then stays at plateaus for $6 \lesssim \D \lesssim 10$.
After that, it increases again to reach the maximal volume-law entanglement, $S_i \simeq \min (i, n - i)$ at $\D \gtrsim 13$.
Those intermediate plateaus resemble the pre-thermalization regime~\cite{Berges2004} in nonequilibrium dynamics.
On the other hand, for the $\fSimpp$ sequence, the intermediate plateaus are almost invisible and $S_i$ quickly approaches the volume-law limit.
It shows that the $\fSimpp$ gates generate more entanglement than others, which is consistent with our PEPS results shown in the main text.
Such gate dependence can also be seen in the entanglement spectra, shown in Figs.~\ref{fig:degeneracy}(a)--\ref{fig:degeneracy}(c).
The $\fSimpp$ sequence generates Schmidt coefficients larger than the other two sequences, rapidly reaching a flat entanglement spectrum.

On the other hand, the diagonal elements $\{ \sigma^{(i,j)}_{\underline{1}} \}$ of the $\Lambda$ tensors exhibit somewhat different gate dependence [Figs.~\ref{fig:degeneracy}(d)--\ref{fig:degeneracy}(f)], reflecting the degeneracy of the OSCs of two-qubit gates (cf.~\Sec{sec:Fex vs Fapx}).
The $\Lambda$ tensor elements show degeneracies (i.e., plateaus) for the $\CZ$ and $\fSimpp$ sequences, while in the 2HR case, they are nondegenerate.
Such degeneracies explain the steplike wiggles in $\mc{F} (\D)$ for the former two sequences, shown in \Fig{fig:exact_fid} and discussed in \Sec{sec:Fex vs Fapx}.
For a given $\D$, the $\Lambda$ tensor elements for the $\fSimpp$ sequence show a more pronounced tail than others, which is consistent with the above observation that it generates more entanglement.

%

\end{document}